\documentclass[aps, reprint]{revtex4-2}
\usepackage[utf8]{inputenc} 
\usepackage[T1]{fontenc} 
\usepackage{lmodern} 
\usepackage{bbm}
\usepackage{graphicx} 
\graphicspath{{./figures/}}
\usepackage[caption=false]{subfig}

\usepackage{orcidlink}

\usepackage{siunitx}
\DeclareSIUnit{\gauss}{G}
\DeclareSIUnit{\Mole}{M}

\usepackage{csquotes} 
\usepackage{xcolor} 

\usepackage{mathtools} 
\usepackage{bm} 
\usepackage{txfonts} 
\DeclareMathAlphabet{\mathoms}{OMS}{cmsy}{m}{n}

\usepackage{tikz}

\usepackage{lipsum}

\usepackage{xspace} 

\begin{document}

\preprint{AIP/123-QED}

\title[Machine learning IVPs]{Solving physics-based initial value problems with unsupervised machine learning}

\def\jqc{Joint Quantum Centre (JQC) Durham--Newcastle, Department of Physics, Durham University, Durham DH1 3LE, United Kingdom}

\author{Jack Griffiths \orcidlink{0000-0001-7794-6687} }\email{me@jackg.co}\affiliation{\jqc}
\author{Steven A. Wrathmall \orcidlink{0000-0003-1770-9721} }\email{s.a.wrathmall@durham.ac.uk}\affiliation{\jqc}
\author{Simon A. Gardiner \orcidlink{0000-0001-5939-4612} }\email{s.a.gardiner@durham.ac.uk}\affiliation{\jqc}

\date{\today}

\begin{abstract}
Initial value problems --- a system of ordinary differential equations and corresponding initial conditions --- can be used to describe many physical phenomena including those arise in classical mechanics. We have developed a novel approach to solve physics-based initial value problems using unsupervised machine learning. We propose a deep learning framework that models the dynamics of a variety of mechanical systems through neural networks. Our framework is flexible, allowing us to solve non-linear, coupled, and chaotic dynamical systems. We demonstrate the effectiveness of our approach on systems including a free particle, a particle in a gravitational field, a classical pendulum, and the Hénon--Heiles system (a pair of coupled harmonic oscillators with a non-linear perturbation, used in celestial mechanics). Our results show that deep neural networks can successfully approximate solutions to these problems, producing trajectories which conserve physical properties such as energy and those with stationary action. We note that probabilistic activation functions, as defined in this paper, are required to learn any solutions of initial value problems in their strictest sense, and we introduce coupled neural networks to learn solutions of coupled systems.

\end{abstract}


\maketitle

\section{Introduction}
Machine learning models embedded with physical constraints, equations, data, or other physical principles (including variational principles such as the Rayleigh--Ritz method or Hamilton's principle, which is what we present in the current work) offer a powerful framework for ensuring physically consistent solutions for differential equations \cite{lagaris, karniadakis, han, sirignano, lu, xu, zhang} or discovering new physics \cite{bongard, schmidt, brunton, raissi2, meng, chen}. Incorporating informed priors in machine learning models allows effective handling of sparse data, improved generalisation, and offer deeper insights into the underlying mechanisms of the systems they are designed to analyse.

Machine learning is an optimisation procedure. Machine learning is \emph{physics-informed} if physical data or principles are present in i) the training data, ii) the underlying neural network architecture, iii) the cost function and iv) the optimisation algorithms. Training data may include data from physical experiments or simulations \cite{greydanus, cranmer}. Training data may be augmented \cite{lecun3} by applying transformations (e.g., translations, scalings or rotations) in line with known symmetries of the system, without having to generate new data. The network architecture may include structures which obey physical laws. Ling, Kurzawski and Templeton \cite{ling} predict turbulent fluid flows using a deep neural network with an auxiliary network which ensures invariance to Galilean transformations (which one would expect from such classical fluid problems). 

Regularisation terms may be added to the cost function to constrain the neural network output based on, for example, the underlying differential equation, whether the solution should be divergence free \cite{wang-yu}, or any other mathematical description of the data \cite{raissi}. The cost function therefore may consist of a weighted combination of: i) the differential equation (or other equations which constrain the solution), ii) the initial conditions, iii) the boundary conditions (where appropriate), and iv) sparse data sampled from throughout the problem domain (the training data). Other problem-specific conditions may also be included in the cost function. Xu, \emph{et al.} \cite{xu}, solve Fokker--Planck equations with machine learning methods, which additionally \emph{requires} constraining the normalisation of their solutions (which are probability distributions). The training data may be sparse, as this approach effectively interpolates between data points by enforcing the constraint given by the differential equation.

We stress through our own experiments and through earlier research of Vapnik \cite{vapnik, vapnik2} that one should choose the simplest cost function possible to avoid overfitting neural networks or --- in some cases --- complete failure of their training. One may know the differential equation, initial conditions, Hamiltonian \cite{greydanus}, and Lagrangian \cite{cranmer}, but one should not, in general, include all of these terms in the cost function.

In the present work, we describe a framework to solve initial value problems in the strictest sense, that is without any constraints which are not needed mathematically. We do not include any extraneous regularisation terms as is common in typical physics-informed neural networks \cite{lagaris, karniadakis, han, sirignano, lu, xu}. Our framework is an implementation of Hamilton's principle through machine learning methods; for all dynamical problems we consider, the only constraints are the underlying differential equation and its associated initial conditions. There is no training data --- this is \emph{unsupervised} learning. Our discussion starts with the most minimally conceivable neural network to describe free particle dynamics --- this allows us to set notation and introduce our neural network representation of initial value problems. We consider linear approximations of a particle in a gravitational field and in a harmonic oscillator. We briefly discuss observed deviations from the theoretical results of the harmonic oscillator. Guided by the need to increase the representational capacity of our neural networks, we introduce deep learning techniques to solve the classical, arbitrary-angle pendulum. We introduce coupled neural networks and optimisation routines, using the Hénon--Heiles system \cite{henon} (akin to three-body dynamics \cite{bastos}) as an example. We finally make comments relating the machine learning trajectories back to Hamilton's principle --- our solutions are eventually consistent with the principle of stationary action, despite some peculiarities in how the action evolves during training.

\section{Statement of the problem}
\subsection{Statement of the physical problem}
Consider a mechanical system characterised by $m$ generalised coordinates $\{q_i(t)\}$, $i\in\{1,\cdots,m\}$, and described by the Lagrangian $L$. Application of Hamilton's principle leads to the Euler--Lagrange equations
\begin{equation}
  \frac{\mathrm{d}}{\mathrm{d}t}\frac{\partial L}{\partial \dot{q}_i} - \frac{\partial L}{\partial {q}_i} = 0.
\end{equation}
We chose to consider only Lagrangians of the form $L=T-V$, where
\begin{equation}
  T = \frac{1}{2}\sum_{i=1}^m \mu_i\dot{q}_i(t)^2
\end{equation}
is the kinetic energy and $V\equiv V(q_1(t), q_2(t), \cdots, q_m(t))$ is a potential energy. We consider only potential energies that are explicit functions of the generalised coordinates, which excludes time-dependent potentials and the Lorentz force. The Euler--Lagrange equations lead to equations of motion of the form
\begin{equation}
  \mu_i\ddot{q}_i(t) - F_i(q_1(t), q_2(t), \cdots, q_m(t)) = 0,
\end{equation}
where $\mu_i$ are generalised coefficients --- these may be associated with a mass or moment of inertia, depending upon the dynamical system --- and $q_i$ are the generalised coordinates --- which may be associated with positions or angles or other suitable coordinates that uniquely define the configuration of the system. Their dimensions will depend on the nature of each coordinate (e.g., an angle coordinate will be dimensionless, whereas a length coordinate will have dimensions of length). We can associate
\begin{equation}
  F_i(q_1(t), q_2(t), \cdots, q_m(t)) = -\frac{\partial V(q_1(t), q_2(t), \cdots, q_m(t))}{\partial q_i(t)},
  \label{ch3:force}
\end{equation}
with a generalised force. In order to eliminate the explicit dependence upon $\mu_i$, let the rescaled generalised force (which may have units of acceleration or angular acceleration, as appropriate) be $f_i = F_i/\mu_i$. This leads to the equations of motion
\begin{equation}
  \ddot{q}_i(t) - f_i(q_1(t), q_2(t), \cdots, q_m(t)) = 0,
  \label{eq:ivp}
\end{equation}
where we choose to cast the equations of motion in such a way that they equal zero. This simplifies the subsequent development of the machine learning problem in section \ref{ml_prob}.

The domain of the problem is $t\in[0, T]$ (the time origin is set to zero without loss of generality). The problem is subject to $2m$ initial conditions, where each generalised coordinate is associated with an initial position $s_{i,0}$ and velocity $v_{i,0}$.

\subsection{Statement of the machine learning problem}\label{ml_prob}
We will denote the neural network's representation of the solution by $\mathbf{q}_i(\mathbf{t}) = \left( q_{i,0}, q_{i,1}, \cdots, q_{i,n}, \cdots, q_{i,N_T} \right)$. Each element of the neural representation, $q_{i,n}$, is associated with the continuous solution at a particular time $t_n=n\Delta t$, i.e., $q_{i,n} \approx q_i(n\Delta t)$, where $\Delta t = T/N_T$ is the time step. We are seeking discretised vector representations $\{\mathbf{q}_i(\mathbf{t})\}$ of the continuous functions $\{q_i(t)\}$ over the temporal domain $\mathbf{t}=(0, t_1, \cdots, t_n, \cdots, t_{N_T})$.

The machine learning problem is to find an optimised neural network representation which approximates the solution of the initial value problem by minimising an appropriate cost metric. Our cost metric will be a dimensionless function (to not be dependent upon any particular system of units). We will associate all timescales with a characteristic time $\tau$ and all length scales with a characteristic length $\lambda_i$. This implies a dimensionless generalised position coordinate $\tilde{q}_i=q_i/\lambda_i$ and a dimensionless rescaled generalised force $\tilde{f}_i = f_i\tau^2/\lambda_i$. Furthermore, the dimensionless parameters $\alpha_i, \beta_i$ and $\gamma_i$ may be introduced to adjust the relative weighting of each term in the cost function. Dropping all tildes, the dimensionless cost function we wish to minimise is
\begin{equation}
  \begin{aligned}
    \mathcal{C} &= \sum_{i=1}^m \frac{1}{\lambda_i^2} \Big[  \alpha_i\tau^4 \left\| \mathbf{\ddot{q}}_i(\mathbf{t}) - \mathbf{f}_i\left(\mathbf{q}_1(\mathbf{t}), \mathbf{q}_2(\mathbf{t}), \cdots,  \mathbf{q}_m(\mathbf{t}) \right) \right\|^2 + \\
    &\quad\quad\quad\quad + \beta_i \tau^2 \left(\dot{q}_{i,0} - v_{i,0}\right)^2 + \gamma_i \left(q_{i,0} - s_{i,0}\right)^2 \Big]
  \end{aligned}
  \label{ch3:pinn_ivp_loss}
\end{equation}
where $\mathcal{C}$ produces, in general, a positive, real and scalar output. We define, in general, the notation
\begin{equation}
   \|\mathbf{g}(\mathbf{t})\|^2 \equiv \frac{\Delta t}{\tau} \sum_{n=0}^{N_T} g(t_n)^2 = \frac{T}{\tau N_T} \sum_{n=0}^{N_T} g(t_n)^2,
\end{equation}
such that
\begin{equation}
  \lim_{\Delta t \to 0} \|\mathbf{g}(\mathbf{t})\|^2 = \frac{1}{\tau} \int_{0}^T \mathrm{d}t\, g(t)^2.
\end{equation}

Scaling the generalised coordinates, $q_i$, and time, $t$, to be expressed in terms of $\lambda_i$ and $\tau$, respectively, ensures a dimensionless form; this corresponds to setting $\lambda_i=\tau=1$, which we do in every example in this paper. Furthermore, we always set $\alpha_i=\beta_i=\gamma_i=1$. Under these assumptions, the cost function takes the form
\begin{equation}
  \begin{aligned}
    \mathcal{C} &= \sum_{i=1}^m \Big[  \left\| \mathbf{\ddot{q}}_i(\mathbf{t}) - \mathbf{f}_i\left(\mathbf{q}_1(\mathbf{t}), \mathbf{q}_2(\mathbf{t}), \cdots,  \mathbf{q}_m(\mathbf{t}) \right) \right\|^2 + \\
  &\quad\quad\quad + \left(\dot{q}_{i,0} - v_{i,0}\right)^2 + \left(q_{i,0} - s_{i,0}\right)^2 \Big]
  \end{aligned}
  \label{ch3:pinn_ivp_loss_simple}
\end{equation}
The first line is the residual of the differential equation (i.e., the differential equation should equal zero). The second line constrains $\dot{q}_i(0)$ and $q_i(0)$ to not deviate significantly from the initial velocity, $v_{i,0}$, and the initial position, $s_{i,0}$. The cost function, in principle, fully specifies the initial value problem, and when the differential equation and initial conditions are satisfied, the neural network solution should tend towards the expected solution of the initial value problem.

\section{Linear neural network representation of initial value problems}
\subsection{Overview}
In order to set notation, we first explore free particle dynamics by considering the simplest conceivable neural network for modelling linear solutions to initial value problems, as shown in Fig. (\ref{ch3:ivp_constant_velocity_nn}). This neural network can be used to learn free particle dynamics and can behave as a linear approximant for a particle in a gravitational field. We briefly explore using the linear neural network to model the harmonic oscillator.
\subsection{Linear neural network for a free particle}

\subsubsection{Statement of the equation of motion and cost function}
Consider a system with the free particle Lagrangian
\begin{equation}
  L = \frac{1}{2}\mu\dot{q}(t)^2.
\end{equation}
Solving the Euler--Lagrange equations leads to the equation of motion
\begin{equation}
  \ddot{q}(t)=0,
  \label{ch3:ivp_constant_velocity}
\end{equation}
subject to the initial position $s_0$ and initial velocity $v_0$. Equation (\ref{ch3:ivp_constant_velocity}) has solutions $q(t)=v_0 t + s_0.$

\begin{figure}[t!]
  \centering
  \includegraphics[width=\linewidth]{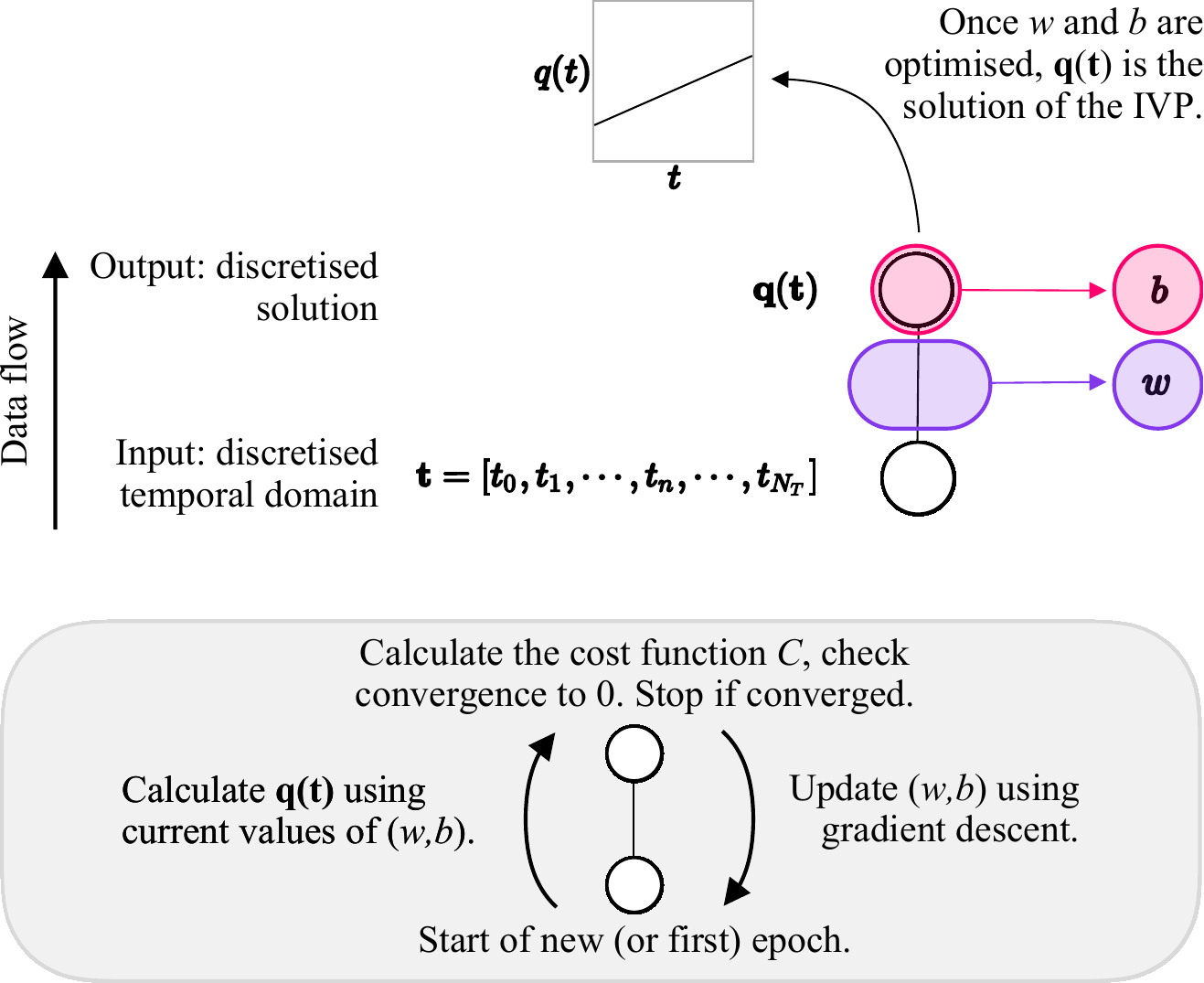}
  \caption[Neural network to solve the constant velocity problem]{A neural network to solve the differential equation (\ref{ch3:ivp_constant_velocity}), as represented by Eq. (\ref{eq:constant_velocity_discretised_soln}). The network contains two neurons: i) an input neuron, which represents the discretised temporal domain, and ii) an output neuron, which represents the discretised solution to the initial value problem. The network consists of two parameters: a weight between the input and output vector and a bias associated with the output parameter.}
  \label{ch3:ivp_constant_velocity_nn}
\end{figure}

For this dynamical system, we seek neural solutions
\begin{equation}
  \mathbf{q}(\mathbf{t}) = w\mathbf{t} + b\mathbbm{1},
  \label{eq:constant_velocity_discretised_soln}
\end{equation}
where $\mathbbm{1}\in\mathbb{R}^{N_T}$ is a vector of ones.

Figure \ref{ch3:ivp_constant_velocity_nn} represents the discretised solution in Eq. (\ref{eq:constant_velocity_discretised_soln}) as a neural network. Neural networks are composed of nodes connected by edges, forming a graph where each node is referred to as a neuron. In this paper, neurons are always vectors --- this choice means that there are fewer weights and biases to learn and that a direct equivalency of Eq. (\ref{eq:constant_velocity_discretised_soln}) with linear regression may be observed. Every neuron-neuron connection is associated with a weight, and every neuron (in all layers except the input) is associated with a bias. The free particle system is described entirely by the simplest conceivable neural network: it contains no hidden layers or activation functions. We define the zeroth layer neuron value as $\mathbf{t}$ and the output layer neuron as $\mathbf{q}(\mathbf{t})$. The identification of $v_0$ with the weight $w$ and $s_0$ with the bias is trivial, i.e., $q(0) = b$, and $\dot{q}(t)=\dot{q}(0)=w$. We can, therefore, describe the cost function as a function of the weight and bias,
\begin{equation}
  \begin{aligned}
    \mathcal{C}(w,b) &= \| f(w\mathbf{t} - b\mathbbm{1}) \|^2 + (w-v_0)^2 + (b-s_0)^2 \\
    &= \frac{T}{N_T} \sum_{n=0}^{N_T} f(wt_n+b)^2 + (w-v_0)^2 + (b-s_0)^2.
  \end{aligned}
\end{equation}
Given that $\ddot{q}=0 \implies f(\cdot)=0$, the cost function reduces to
\begin{equation}
  \mathcal{C}(w,b) = (w-v_0)^2 + (b-s_0)^2,
  \label{ch3:constant_velocity_loss}
\end{equation}
which is minimised by $w=v_0$ and $b=s_0$. 

\subsubsection{Initialisation}\label{linear_net_init}
\begin{figure}[ht!]
  \centering
  \includegraphics[width=\linewidth]{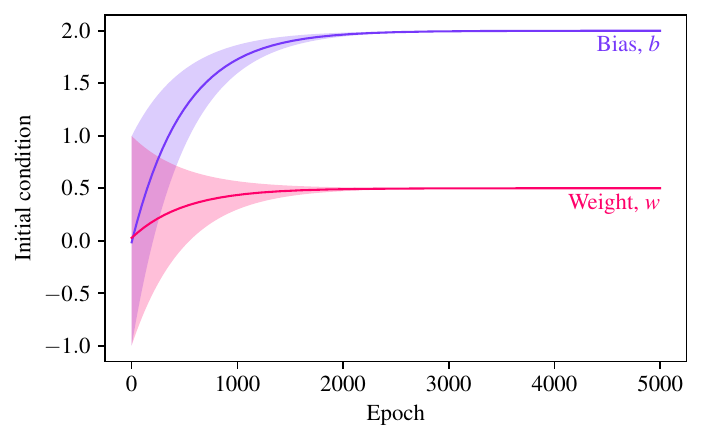}%
  \caption[Neural solution and convergence of parameters (weight and bias) for the constant velocity problem]{\emph{Free particle:} The convergence of the weight and bias for an initial value problem describing a free particle with $s_0=2.0$ and $v_0=0.5$ for 250 instances of the initialisation procedure. The weight and bias are initialised randomly according to Eq. (\ref{ch3:constant_velocity_initialisation}) and tend towards their expected values after sufficient iterations of an optimisation algorithm have passed. After around 2,500 epochs, the training may be stopped as an approximate solution has been obtained. The shaded region indicates the maximum and minimum value of both parameters at a given epoch, and the line indicates the average value of the parameters throughout (which is an arithmetic mean for the underlying uniform distribution).}
  \label{ch3:ivp_constant_velocity_parameter_convergence}
\end{figure}

All parameters in any machine learning model are initialised randomly. This initialisation has no physical interpretation; it is part of the computational methodology, as explored in appendix \ref{ch1:initialisation}. The weight and bias for the neural network are sampled from the uniform distribution \cite{he}
\begin{equation}
  w, b \sim U\left( -1, 1 \right).
  \label{ch3:constant_velocity_initialisation}
\end{equation}
The range of initial values of the weight and bias and their convergence to the expected parameters, $v_{i,0}$ and $s_{i,0}$, for 250 iterations of the training procedure is shown in Fig. (\ref{ch3:ivp_constant_velocity_parameter_convergence}).

\subsubsection{Learning the optimal parameters using gradient descent}\label{gradient_descent}
Optimisation of a two parameter model can be achieved by gradient descent (usually attributed to Cauchy). Let the model parameters be denoted by the tuple $(w, b)$. We wish to adjust the weight and bias by some $\Delta w$ and $\Delta b$, respectively, where the adjustment is assumed to be small to avoid overshooting the global minima. Let the adjustments be described by the tuple $(\Delta w, \Delta b)$. Ensuring a descent in the direction of the negative gradient, the result of the first order expansion of the cost function about $(\Delta w, \Delta b)$ is
\begin{equation}
  (\Delta w, \Delta b) = -\eta \left( \frac{\partial\mathcal{C}(w,b)}{\partial w}, \frac{\partial\mathcal{C}(w,b)}{\partial b} \right),
  \label{eq:gradient_descent_const_vel}
\end{equation}
where $\eta>0$ is a fixed global hyperparameter (a parameter which describes a neural network) known as the step size or learning rate. Equation (\ref{eq:gradient_descent_const_vel}) is the epoch-to-epoch update formula for the weight and the bias.

\begin{figure}
  \centering
  \includegraphics[width=\linewidth]{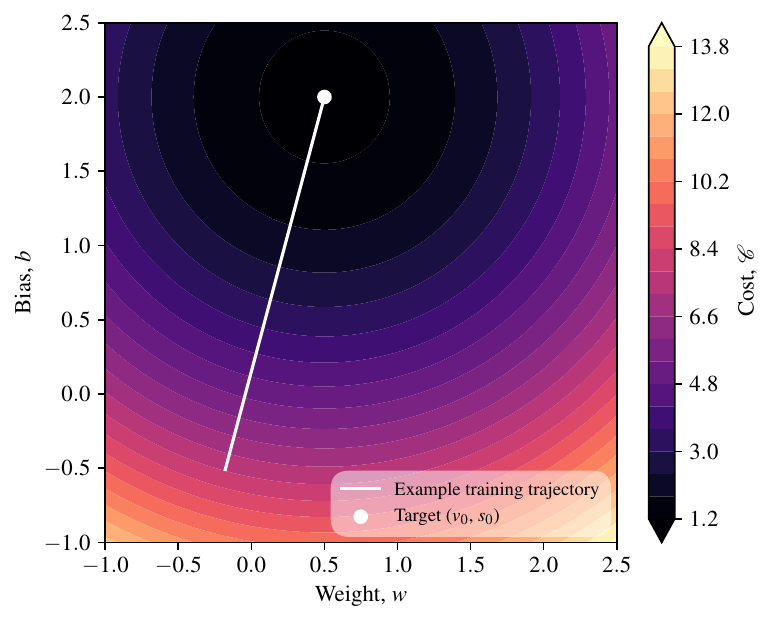}
  \caption[Gradient descent in a two-parameter cost landscape]{The gradient descent/update procedure for the weights and biases towards the target parameters $v_0$ and $s_0$ are shown, according to Eq. (\ref{eq:gradient_descent_const_vel}). The contours represent the values of the cost function in Eq. (\ref{ch3:constant_velocity_loss}).}
  \label{ch3:ivp_constant_velocity_descent}
\end{figure}

From the original definition of the cost function in Eq. (\ref{ch3:constant_velocity_loss}),
\begin{equation}
 \frac{\partial C(w,b)}{\partial w}=2(w-v_0)\quad\text{and}\quad \frac{\partial C(w,b)}{\partial b}=2(b-s_0).
\end{equation}
Equation (\ref{eq:gradient_descent_const_vel}) describes the gradient descent throughout the parameter space. One set of parameter updates is associated with one epoch (one complete forward and backwards pass through the neural network). After a sufficient number of epochs have been completed, the weight and bias should tend toward the dimensionless initial velocity and initial position, respectively. Figure \ref{ch3:ivp_constant_velocity_descent} shows the gradient descent of the weight and biases throughout the parameter space, updating according to Eq. (\ref{eq:gradient_descent_const_vel}). 

\begin{figure}[ht!]
  \centering
  \includegraphics[width=\linewidth]{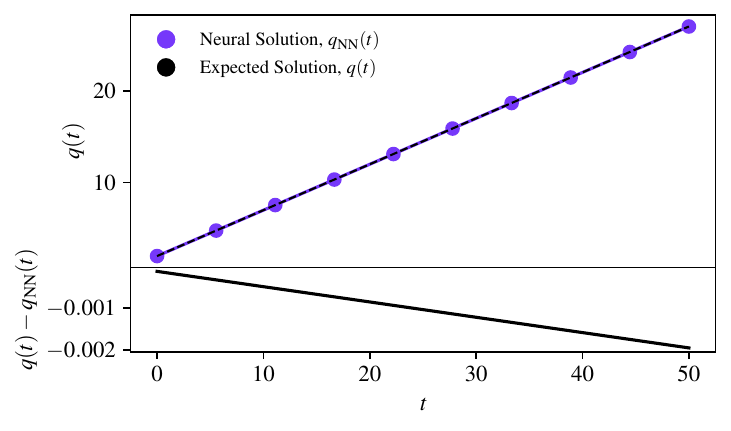}
  \caption[Neural solution and convergence of parameters (weight and bias) for the constant velocity problem]{\emph{Free particle:} The expected solution, the neural solution, and the difference (residual) of the two are shown.}
  \label{ch3:ivp_constant_velocity_solutions}
\end{figure}
Figure \ref{ch3:ivp_constant_velocity_solutions} demonstrates the neural solution given by Eq. (\ref{eq:constant_velocity_discretised_soln}) for the learnt parameters $w$ and $b$. Given that the values of $s_0$ and $v_0$ obtained by the neural network are approximations, the solution they describe will always diverge from the expected solution after sufficient time, as shown by the residual plot.

\subsection{Linear neural network as a linear approximant for a particle in a gravitational field}

\begin{figure}
  \centering
  \subfloat[\label{subfig:a}]{%
  \includegraphics[width=\linewidth]{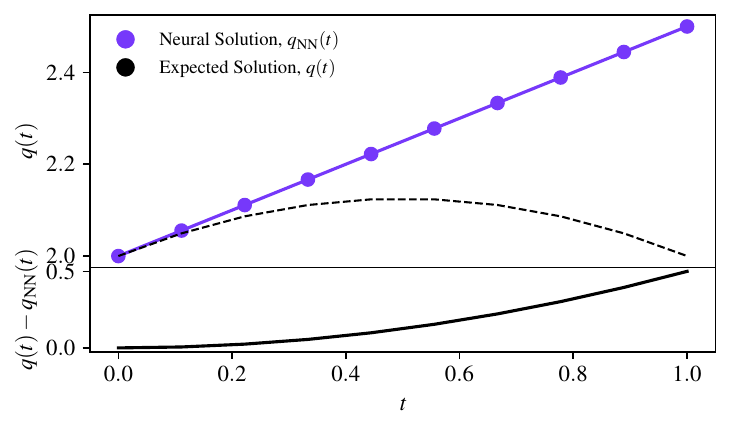}%
  }\\
  \subfloat[\label{subfig:b}]{%
  \includegraphics[width=\linewidth]{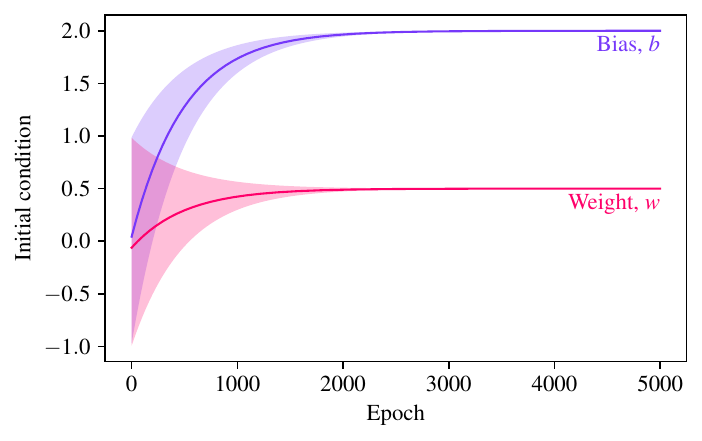}%
  }
  \caption[Linear neural network to approximate early temporal solutions of non-linear solutions]{(a) The linear approximant to Eq. (\ref{eq:constant_velocity_discretised_soln}) for a particle in a gravitational field obtained by machine learning, alongside the expected particle trajectory. Since we are trying to fit a linear function to a quadratic function, the residuals show an underlying quadratic pattern as time evolves. (b) The average convergence of the bias and weight for 250 iterations of the training algorithm is again shown. The shaded region indicates the maximum and minimum value of both parameters at a given epoch, and the line indicates the average value of the parameters throughout (which is an arithmetic mean for the underlying uniform distribution).}
  \label{ch3:ivp_gravitational_field_solution_params}
\end{figure}
This framework additionally serves as a linear approximant for functions which exhibit non-linear solutions. Consider a particle in a uniform gravitational field given by the dimensionless equation of motion
\begin{equation}
  \ddot{q}(t)+1 = 0,
\end{equation}
which --- in the linear neural network framework --- continues to have solutions of the form in Eq. (\ref{eq:constant_velocity_discretised_soln}). The cost function is again a function of the weight and bias, with a non-zero forcing term
\begin{equation}
  \begin{aligned}
    \mathcal{C}(w,b) &= \frac{T}{N_T}\sum_{n=0}^{N_T} 1^2 + (w-v_0)^2 + (b-s_0)^2 \\
    &= \frac{T}{N_T} (N_T+1) + (w-v_0)^2 + (b-s_0)^2 \\
    &\approx T + (w-v_0)^2 + (b-s_0)^2 \\
  \end{aligned}
\end{equation}
as $N_T\to\infty$. The cost function is again minimised by $w$ and $b$ but increases without bound as $T\to\infty$. Figure \ref{ch3:ivp_gravitational_field_solution_params} demonstrates the linear neural network's ability to perform as a linear approximant for non-linear solutions.

\subsection{Linear neural network as a linear approximant for a harmonic oscillator}
Consider now the harmonic oscillator given by the equation of motion
\begin{equation}
  \ddot{q}(t) + q(t) = 0.
\end{equation}
Letting $f(t)=-q(t)=-wt-b$, the cost function is
\begin{equation}
  \begin{aligned}
    \mathcal{C}(w,b) &= \frac{T}{N_T}\sum_{n=0}^{N_T} \left[ w(t_n)+b \right]^2 + (w-v_0)^2 + (b-s_0)^2 \\
    &\approx \int_0^{T}\mathrm{d}t\, \{(wt+b)^2\} + (w-v_0)^2 + (b-s_0)^2\\
    &= \frac{w^2}{3}(T-t_0)^2 + wb(T-t_0)^2 + b^2(T-t_0)+ \\
    &\quad + (w-v_0)^2 + (b-s_0)^2,
  \end{aligned}
\end{equation}
which is minimised by
\begin{equation}
  \frac{\partial \mathcal{C}}{\partial w} = \frac{2}{3} w(T-t_0)^3 +  b(T-t_0)^2 + 2(w-v_0)=0,
  \label{ch5:eq:ho_linear_approx_weight}
\end{equation}
and
\begin{equation}
  \frac{\partial \mathcal{C}}{\partial b} =  w(T-t_0)^2 + 2 b(T-t_0) + 2(b-s_0)=0.
  \label{ch5:eq:ho_linear_approx_bias}
\end{equation}
Solving Eqs. (\ref{ch5:eq:ho_linear_approx_weight}) and (\ref{ch5:eq:ho_linear_approx_bias}) simultaneously gives the weight and bias in terms of the initial conditions $v_0$ and $s_0$ only,
\begin{equation}
  w = \frac{-6T^2 s_0 + 12T v_0 + 12 v_0}{T^4 + 12T+ 4T^3 + 12},
  \label{ch5:eq:ho_approx_weight}
\end{equation}
and
\begin{equation}
  b = \frac{4T^3 s_0 -6 T^2 v_0 + 12 s_0}{T^4 + 12T + 4T^3 + 12}.
  \label{ch5:eq:ho_approx_bias}
\end{equation}
It follows that, as $T\to 0$, $b=s_0$ and $w=v_0$. Furthermore, as $T\to\infty$, $b\to0$ and $w\to 0$. That is to say, when probing long dynamical regimes as $T$ grows larger, the linear approximant suggests that the learnt weight and bias should always tend towards zero regardless of the true values of the initial conditions. We did not observe this, and we did not explore this further. This may suggest the need for increasing the representational capacity of the neural network.

\begin{figure}
  \centering
  \includegraphics[width=\linewidth]{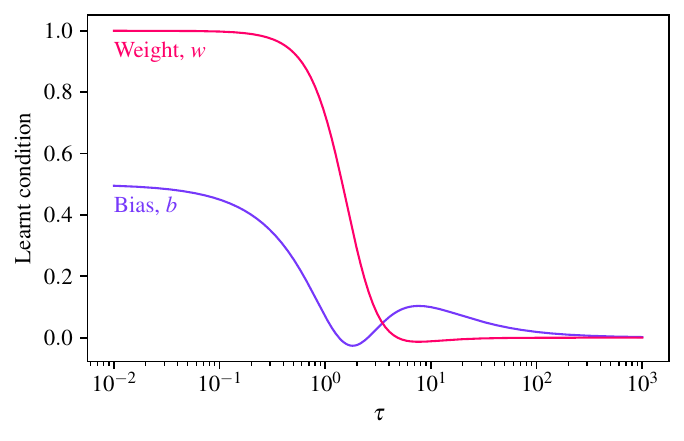}
  \caption{The theoretical values of the weight (Eq. (\ref{ch5:eq:ho_approx_weight})) and bias (Eq. (\ref{ch5:eq:ho_approx_bias})) describing the linear approximant of the harmonic oscillator are shown. The weight and bias are expected to tend towards zero as the time $T$ increases, regardless of the true values of the initial conditions.}
  \label{ch5:ho_linear_approx_params}
\end{figure}

Figure \ref{ch5:ho_linear_approx_params} shows the values of the learnt conditions as functions of $\tau$ for $(s_0,v_0)=(0.5, 1.0)$.

Learning linear solutions or linear approximations to solutions is both computationally inexpensive and easy to follow mathematically. The cost function in Eq. (\ref{ch3:constant_velocity_loss}) is entirely analogous to linear regression in two dimensions (since $\ddot{q}(t)=0$, we are only finding two optimal parameters).

\section{Non-linear neural network representation of initial value problems}
\subsection{Overview}
We will now explore two initial value problems with non-linear solutions: the arbitrary angle pendulum in section \ref{arb_angle} and the Hénon--Heiles system in section \ref{henon}. To model non-linear behaviour, we will extend the already introduced notation to consider a deep learning framework, as shown in Fig. (\ref{ch3:ivp_nonlinear_nn}). Deep learning increases the representational capacity of the neural network by adding more layers (and, in turn, more network parameters) and introduces non-linear (activation) functions. Our work on the Hénon--Heiles system necessitates the introduction of a coupled neural network and optimisation scheme for its solutions.
\subsection{Classical pendulum}\label{arb_angle}
\subsubsection{Statement of the equation of motion and cost function}
Consider a one-dimensional pendulum with kinetic energy
\begin{equation}
  T(\dot{\varphi}(t)) = \frac{1}{2}\mu\ell^2\dot{\varphi}(t)^2,
\end{equation}
where $\mu$ is the mass of the pendulum bob, $\ell$ is the length of the pendulum, and $\dot{\varphi}(t)$ is the angular velocity of the pendulum, associated with the angle coordinate $\varphi(t)$. The potential energy is $V(\varphi(t)) = \mu g\ell[1-\cos\varphi(t)]$, where $g$ is the acceleration due to gravity. The Lagrangian is therefore
\begin{equation}
  L(\varphi(t)) = \frac{1}{2}\mu\ell^2\dot{\varphi}(t)^2 - \mu g\ell[1-\cos\varphi(t)].
\end{equation}
Introducing the characteristic time $\tau = \sqrt{\ell/g}$, and solving the Euler--Lagrange equation, the dimensionless equation of motion is $\ddot{\varphi}(t) + \sin\varphi(t) = 0,$ subject to the pendulum's initial angular velocity $v_{0}$ and initial angle $s_{0}$.

The cost function for this problem is
\begin{equation}
  \begin{aligned}
    \mathcal{C} &= \left\| \pmb{\ddot{\varphi}}(\mathbf{t}) - \sin(\pmb{\varphi}(\mathbf{t})) \right\|^2 + ({\dot{\varphi}}_0 - v_{0})^2 + ({\varphi}_0 - s_{0})^2,
  \end{aligned}
  \label{ch3:pendulum_cost}
\end{equation}
where $\dot{\varphi}_0$ and $\varphi_0$ are the neural network's current values of the initial angular velocity and initial position. The cost function is minimised by an appropriate set of weights and biases, which we will determine soon.

\subsubsection{Towards deep learning}
\begin{figure}[ht!]
  \centering
  \includegraphics[width=\linewidth]{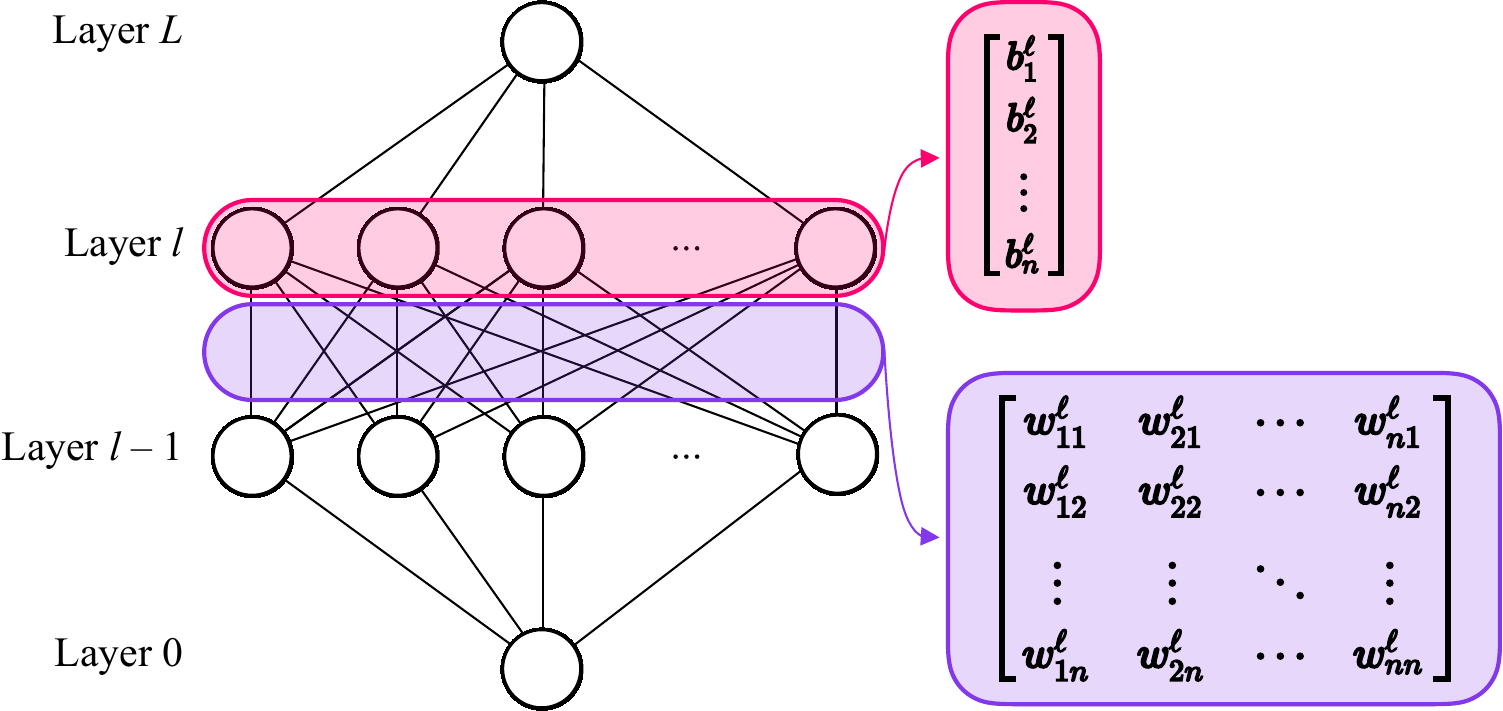}
  \caption[Feedforward, fully-connected deep neural network to find non-linear solutions of dynamical problems]{A fully connected and feedforward deep neural network. We annotate a weights matrix between the $\ell$th and $(\ell-1)$th layer, assuming that both layers contain $n$ neurons. Every connection between every neuron is associated with a weight. We further annotate a bias vector on the $\ell$th layer --- every neuron is associated with a scalar bias. The number of layers and neurons in each layer are hyperparameters which may need to be modified for different dynamical systems.}
  \label{ch3:ivp_nonlinear_nn}
\end{figure}

Unlike the linear neural network, there is no direct physical interpretation of the weights and biases other than in the small region where the solution behaves linearly. For initial value problems where their solutions exhibit non-linear behaviour, we observe that neural networks require of the order of $10^4$ or $10^5$ weights and biases to approximate their solution. We will use a similar architecture to Fig.\,(\ref{ch3:ivp_constant_velocity_nn}) where the input neuron represents the discretised temporal domain, $\mathbf{t}$, and the output neuron represents the discretised solution, $\pmb{\varphi}(\mathbf{t})$. To increase the representational capacity of the neural network, we construct a multi-layer, multi-neuron network as shown in Fig.\,(\ref{ch3:ivp_nonlinear_nn}). Every neuron-neuron connection is associated with a weight parameter and every neuron is associated with a bias parameter. Appendix \ref{ch1:deep_learning} provides further details of the output of a scalar-valued neuron.

The layers between the input and output layers are called hidden layers. Computing successive hidden linear layers would be equivalent to just one linear operation---everything collapses into a linear model so that no non-linear problem can be learnt. To avoid this, we apply a non-linear function $\mathcal{A}$ to every neuron; this is referred to as an activation function, and we exploit the probabilistic interpretation of certain activation functions (see appendix \ref{sec:1_probabilistic_act}) to introduce heuristically a form of model averaging (see appendix \ref{model_avg}) to our solutions. For every neural network describing an initial value problem that we explore in this section, probabilistic activation functions are a fundamental component of the machine learning architecture---non-probabilistic activation functions are not conducive to learning in this context.

\begin{figure}[ht!]
  \centering
  \includegraphics[width=0.9\linewidth]{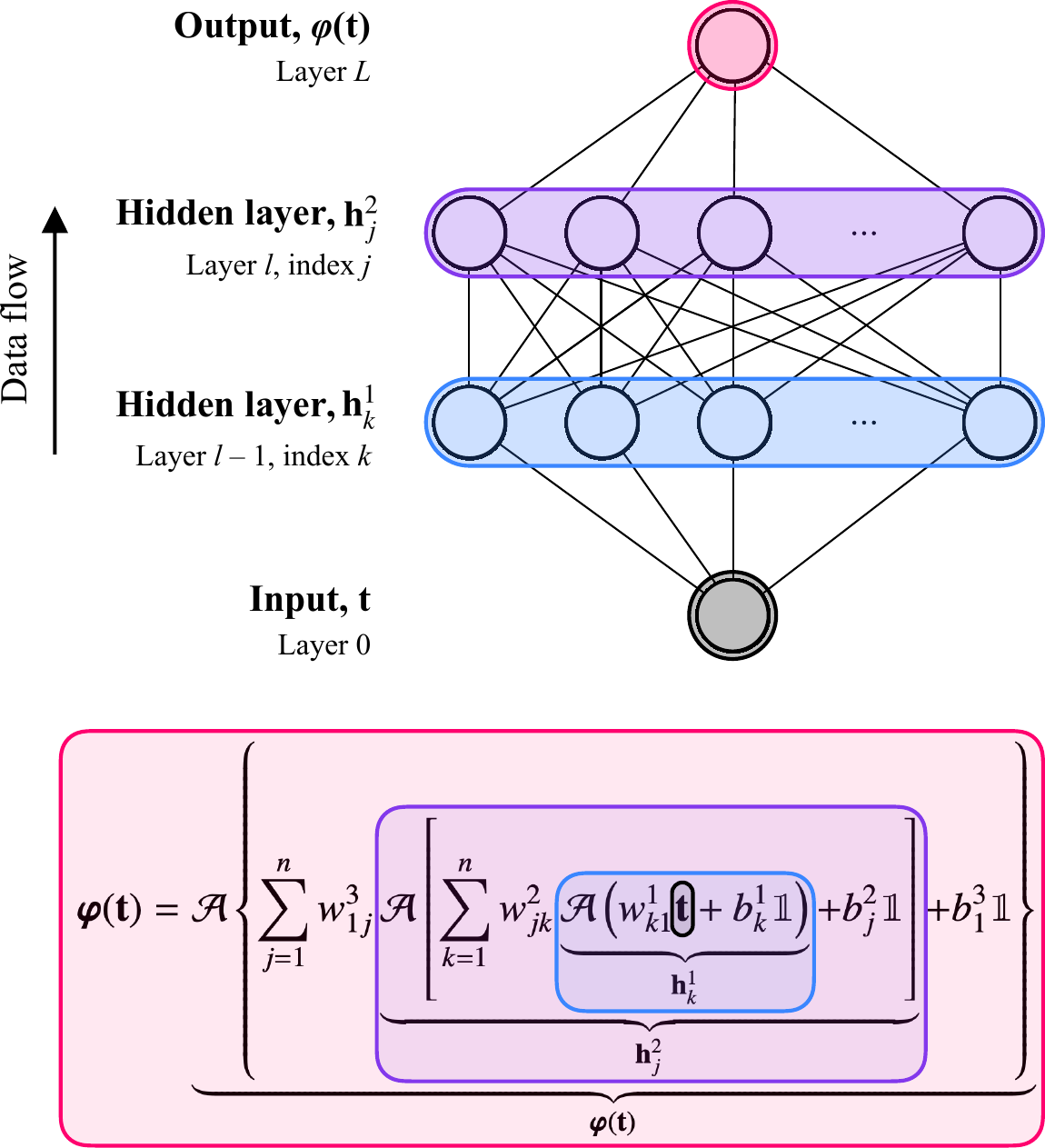}
  \caption[Feedforward, fully-connected deep neural network to find non-linear solutions of dynamical problems]{The neural network in Fig. (\ref{ch3:ivp_nonlinear_nn}) is annotated to show which layers compute Eqs. (\ref{eq:neuron_output_first}--\ref{eq:neuron_output_final}). It demonstrates that the neural network is essentially a composition of many functions. A discretised temporal domain $\mathbf{t}$ with $N_T$ time steps is input to the network. The discretised solution $\pmb\varphi(\mathbf{t})$ is output. Information flows from bottom to top. Every neuron is associated with a tensor of dimensions $[1, N_T, 1]$. Every hidden layer is a tensor of shape $[1, N_T, n]$.}
  \label{fig:arb_angle_pendulum_eq}
\end{figure}

We assume that there are $n$ neurons in each hidden layer (in general, this need not be the case), where each neuron has an index $j\in\{1,n\}$. We further assume that the network has a total of $L$ layers (excluding the input layer, which may be considered the $\ell=0$th layer). Then, the value of a hidden neuron in the first layer is
\begin{equation}
  \begin{aligned}
    \mathbf{h}_k^{1} &= \mathcal{A}(\mathbf{z}_k^{1}),\\
    \mathrm{where}\quad \mathbf{z}_k^{1}&=w_{k1}^{1}\mathbf{t}+b_k^{1} \mathbbm{1},
  \end{aligned}
  \label{eq:neuron_output_first}
\end{equation}
and $\mathbbm{1}\in\mathbb{R}^{N_T}$ is a vector of ones. In the subsequent layer(s), $\ell\in\{2,\cdots,L-1\}$, we similarly have
\begin{equation}
  \begin{aligned}
    \mathbf{h}_j^\ell &= \mathcal{A}(\mathbf{z}_j^{\ell}),\\
    \mathrm{where}\quad \mathbf{z}_j^{\ell}&=\sum_{k=1}^n w_{jk}^{\ell}\mathbf{h}_k^{\ell-1}+b_j^{\ell} \mathbbm{1}.
  \end{aligned}
  \label{eq:neuron_output_arb}
\end{equation}
The final (output) layer, $\ell=L$, of the neural network contains a single neuron with value
\begin{equation}
  \begin{aligned}
    \pmb{\varphi}(\mathbf{t}) &= \mathcal{A}(\mathbf{z}_1^L),\\
    \mathrm{where}\quad \mathbf{z}_1^L&=\sum_{k=1}^n w_{1j}^L\mathbf{h}_j^{L-1}+b^{L}_1 \mathbbm{1}.
  \end{aligned}
  \label{eq:neuron_output_final}
\end{equation}
We find that $L=3$ layer neural networks appear to be sufficient to learn the solutions to initial value problems, including the classical pendulum. Figure \ref{fig:arb_angle_pendulum_eq} shows the complete neural network representation of the classical pendulum in a three-layer architecture (with two hidden layers and one output layer), alongside Eqs. (\ref{eq:neuron_output_first}--\ref{eq:neuron_output_final}).

\subsubsection{Initialisation}
We initialise the weights and biases in the hidden layers from the uniform distribution
\begin{equation}
  w_{jk}^\ell, b_j^\ell \sim U\left(-\sqrt{\frac{3}{n}}, +\sqrt{\frac{3}{n}}\right).
\end{equation}
See appendix (\ref{ch1:initialisation}) for more details.

\subsubsection{Learning the optimal parameters using adaptive moment estimation}
We wish to learn the cost function minimising mapping between the input layer and the output layer. Gradient descent will eventually reach a global minimum deterministically, although it may take a very long time to traverse flat regions of the parameter space where parts of the gradient of the cost function are approximately zero. Instead, in all examples in this section, we use the adaptive moment estimation (Adam) algorithm, which we derive in appendix \ref{ch1:adam}. The gradients of the weights and biases are determined by an algorithm called \emph{backpropagation}. The standard equations of backpropagation for scalar-valued neurons are presented in appendix \ref{backprop_scalar}. We will now derive the equations of backpropagation for vector-valued neurons.

\subsubsection{Backwards propagation of errors with vector-valued neurons}
Forward propagation computes an estimate of the vector $\pmb\varphi(\mathbf{t})$ based on the existing weights and biases. The cost function, $\mathcal{C}$, is then calculated for that particular estimate. We want to determine the \emph{sensitivity} of the cost function with respect to the weights and biases throughout the entire network. The sensitivities of the cost function are given by
\begin{subequations}
  \begin{align}
    \frac{\partial\mathcal{C}}{\partial w_{jk}^\ell} &=  \nabla_{\mathbf{z}_j^\ell}\mathcal{C} \cdot \nabla_{w_{jk}^\ell}\mathbf{z}_j^\ell
    \label{eq:backprop_w}\\
    \frac{\partial\mathcal{C}}{\partial b_{j}^\ell} &=  \nabla_{\mathbf{z}_j^\ell}\mathcal{C} \cdot \nabla_{b_{j}^\ell}\mathbf{z}_j^\ell = \nabla_{\mathbf{z}_j^\ell}\mathcal{C} \cdot \mathbbm{1} = \nabla_{\mathbf{z}_j^\ell}\mathcal{C}.
    \label{eq:backprop_b}
  \end{align}
\end{subequations}
which we obtain by the chain rule. We use $\cdot$ to denote the scalar product for notational simplicity.
Using the definition of the output of a neuron in the first hidden layer in Eq. (\ref{eq:neuron_output_first}) and in an arbitrary layer, $\ell$, in Eq. (\ref{eq:neuron_output_arb}), we can see that the sensitivity of the cost function with respect to the network weights is
\begin{equation}
  \frac{\partial\mathcal{C}}{\partial w_{jk}^\ell} = 
  \begin{cases}
    \nabla_{\mathbf{z}_{j}^{\ell}}\mathcal{C} \cdot \mathbf{h}_j^\ell, & \text{if $\ell\geq 2$,}\\
    \nabla_{\mathbf{z}_{j}^{\ell}}\mathcal{C} \cdot \mathbf{t}, & \text{if $\ell=1$.}
   \end{cases}
  \label{eq:backprop_wrt_weight_vector}
\end{equation}

We seek an equation for $\nabla_{\mathbf{z}_j^\ell}\mathcal{C}$, which is calculated backwards through the network for computational efficiency \cite{griewank}. This is the sensitivity of the cost function with respect to the weighted input to the neuron. In the final layer, this sensitivity is given by
\begin{equation}
  \nabla_{\mathbf{z}_{1}^L}\mathcal{C} = \mathop{\nabla_{\pmb\varphi}\mathcal{C}}\mathop{ \nabla_{\mathbf{z}_{1}^L}{ \pmb\varphi}}.
  \label{eq:weighted_input_sensitivity_vec}
\end{equation}
The sensitivity is, in a linear algebra sense, a Jacobian matrix (of the partial derivatives of the elements of $\pmb\varphi$ with respect to the elements of $\mathbf{z}_{1}^L$) multiplied by a row vector (of the partial derivatives of $\mathcal{C}$ with respect to each element of the solution vector $\pmb\varphi$). Since the activation function is applied element-wise, the Jacobian $\nabla_{\mathbf{z}_{1}^L} \pmb\varphi$ is diagonal, so the sensitivity reduces to an element-wise multiplication, as indicated by the Hadamard product, $\odot$. It therefore follows that
\begin{equation}
  \nabla_{\mathbf{z}_{1}^L}\mathcal{C} = \nabla_{\pmb\varphi}\mathcal{C}\odot(\nabla_{\mathbf{z}_{1}^L} \odot {\pmb\varphi}),
\end{equation}
where the substitution $\pmb\varphi=\mathcal{A}(\mathbf{z}_{1}^L)$ can then be made --- this directly introduces the derivative of the activation function into the backpropagation algorithm.

We then work top to bottom through the neural network (hence, backpropagation) to determine the optimality of each neuron. Let us consider two arbitrary layers in the network, $\ell$ and $\ell-1$, with neuron indices $j$ and $k$, respectively. This preserves the previously defined ordering of indices on, e.g., the weights $w_{jk}^\ell$. For the layer $\ell-1$, the sensitivity of the cost function with respect to the $k$th neuron can be written in terms of all of the sensitivities in the layer after it (this neuron influences all neurons in the subsequent layers $\ell$, $\ell+1$, and so on, in a fully connected neural network). The sensitivity of the output of a neuron in layer $(\ell-1)$ is therefore also determined by the chain rule
\begin{equation}
  \nabla_{\mathbf{z}_{k}^{\ell-1}}\mathcal{C} = \sum_{j=1}^{n_{\ell}} \mathop{\nabla_{\mathbf{z}_{j}^{\ell}}\mathcal{C}} \mathop{\nabla_{\mathbf{z}_{k}^{\ell-1}} \mathbf{z}_j^{\ell}},
  \label{eq:error_in_layer_ell_vec_l-1}
\end{equation}
where each term in the summation is a Jacobian matrix (the matrix of partial derivatives of pre-activations from one layer to the next) multiplied by a vector (the sensitivity in the cost function with respect to a change in the pre-activations in the subsequent layers).

Let the Jacobian matrix $\mathsf{J}^{jk} = \mathop{\nabla_{\mathbf{z}_{k}^{\ell-1}} \mathbf{z}_j^{\ell}}$ with elements 
\begin{equation}
  \mathsf{J}_{mn}^{jk} = \frac{\partial(z_j^{\ell})_m}{\partial(z_k^{\ell-1})_n}.
\end{equation}
By the definition of the pre-activation in Eq. (\ref{eq:neuron_output_arb}), it follows that each element $\mathsf{J}_{mn}^{jk}$ is
\begin{equation}
  \mathsf{J}_{mn}^{jk} =
  \sum_{k'=1}^{n_\ell} w_{jk'}^{\ell} \frac{\partial (\mathcal{A}(({z}_{k'}^{\ell-1})_m))}{\partial (({z}_k^{\ell-1})_n)}, 
  \label{eq:error_in_layer_ell_step1_vec}
\end{equation}
which can only be nonzero if $k'=k$ and $m=n$, such that the function being differentiated is a function of the variable it is being differentiated with respect to.
Therefore, the matrices $\mathsf{J}^{jk}$ are again diagonal, with 
\begin{equation}
  \mathsf{J}_{nn}^{jk} =
  w_{jk}^{\ell} \frac{\partial (\mathcal{A}(({z}_{k}^{\ell-1})_n))}{\partial (({z}_k^{\ell-1})_n)}
  \end{equation}
  and $\mathsf{J}_{m\neq n}^{jk} =0$. Substituting Eq.~(\ref{eq:error_in_layer_ell_step1_vec}) into equation Eq.~(\ref{eq:error_in_layer_ell_vec_l-1}) gives the error in the $(\ell-1)$th layer in terms of the errors in the $\ell$th layer, 
\begin{equation}
  \nabla_{\mathbf{z}_{k}^{\ell-1}}\mathcal{C} =  \sum_{j=1}^{n_{\ell}} w_{jk}^{\ell} \nabla_{\mathbf{z}_{j}^{\ell}}\mathcal{C} \odot \left( \nabla_{\mathbf{z}_k^{\ell-1}} \odot \mathcal{A}(\mathbf{z}_k^{\ell-1}) \right).
\end{equation}
It immediately follows that between the output and the penultimate layer,
\begin{equation}
  \nabla_{\mathbf{z}_{j}^{\ell}}\mathcal{C} = w_{1j}^{L} \nabla_{\mathbf{z}_{1}^{L}}\mathcal{C} \odot \left( \nabla_{\mathbf{z}_j^{\ell}} \odot \mathcal{A}(\mathbf{z}_j^{\ell}) \right).
\end{equation}

\begin{figure}[ht!]
  \centering
  \includegraphics[width=0.9\linewidth]{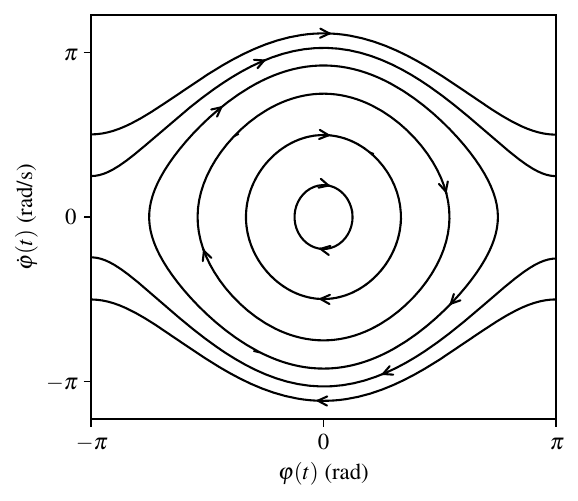}
  \caption[Phase space plot of the neural solutions of the non-linear pendulum]{The phase portrait for four open and four closed trajectories. Small artefacts can be observed (for example, on the closed trajectory with the highest initial angle) since the error in the neural solution increases as time increases. Note that the unstable trajectory at $\varphi(0)=\pi$ and $\dot{\varphi}(0)=0$ does not appear on the phase space portrait --- it only exists at a single point.}
  \label{ch3:nonlinear_pendulum_trajectories}
\end{figure}

\begin{figure*}[ht!]
  \centering
  \includegraphics[width=1.0\textwidth]{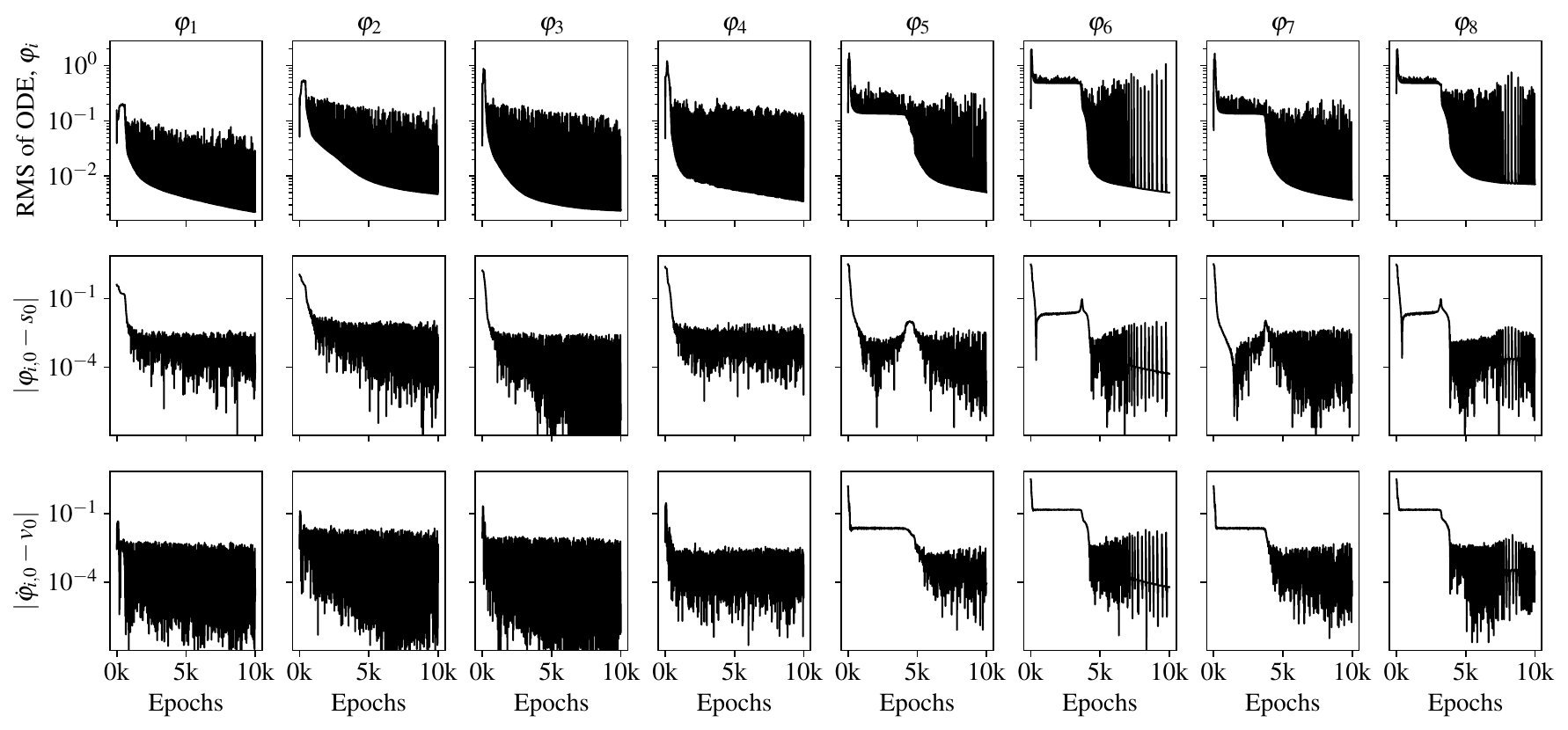}
  \caption{Cost metrics for all trajectories in the classical pendulum that were probed in Fig. (\ref{ch3:nonlinear_pendulum_trajectories}). The closed trajectories have initial angles $\varphi_{1,0}=\pi/8$, $\varphi_{2,0}=\pi/3$, $\varphi_{3,0}=13\pi/24$ and $\varphi_{4,0}=3\pi/4$ (these are four equally spaced angles covering a broad range of dynamical behaviour) and initial angular velocities $\dot{\varphi}_{i=1,\cdots,4} = 0$. The open, positive trajectories have initial angles $\varphi_{5,0}=-\pi$ and $\varphi_{6,0}=-\pi$ and initial angular velocities $\dot{\varphi}_{5,0}=\pi/2$ and $\dot{\varphi}_{6,0}=\pi$. The open, negative trajectories have initial angles $\varphi_{7,0}=\pi$ and $\varphi_{8,0}=\pi$ and initial angular velocities $\dot{\varphi}_{7,0}=-\pi/2$ and $\dot{\varphi}_{8,0}=-\pi$. All cost metrics use a logarithmic scale.
  }
  \label{fig:pendulum_cost_metrics}
\end{figure*}

\subsubsection{Results and discussion}
The exact solution for the pendulum dynamics, for arbitrary initial angle and angular velocity, may be written in terms of Jacobi elliptic functions, but the standard method of solution is via numerical integration schemes. There is a numerical instability when $\varphi(0)=\pi$ and $\dot{\varphi}(0)=0$; the pendulum should remain at $\varphi(t)=\pi$ for all time, but machine precision means that the pendulum will always fall, as the pendulum is never at exactly $\varphi(0)=\pi$.

We sample both open and closed trajectories of the classical pendulum. The closed trajectories have initial positions from $s_{0}=\pi/8$ to $s_{0}=3\pi/4$ --- equally spaced and chosen to cover a broad range of dynamical behaviour --- and zero velocity, $v_{0}=0$. The open trajectories have initial positions $s_{0}=\pm\pi$ and velocities $v_{0}=\pm\pi/2$ or $v_{0}=\pm\pi$ (the open and positive trajectories have negative initial angle and positive initial angular velocity, and the open and negative trajectories have positive initial angle and negative initial angular velocity). All trajectories are shown in Fig. \ref{ch3:nonlinear_pendulum_trajectories}. Figure \ref{fig:pendulum_cost_metrics} shows the various cost metrics for each trajectory. The machine learning modelling estimates the solution of the initial value problem for a given set of weights and biases, and determines the mean square error of the trajectory across the problem domain. In order to plot a quantity with physical dimensions, we take the square root of this estimate; we define the quantity $\sqrt{\left\| \mathbf{\ddot{q}}_i(\mathbf{t}) - \mathbf{f}_i\left(\mathbf{q}_1(\mathbf{t}), \mathbf{q}_2(\mathbf{t}), \cdots,  \mathbf{q}_m(\mathbf{t}) \right) \right\|^2}$ to be the root-mean-square of the differential equation (RMS of the ODE).
Remarkably, we note that the machine learning modelling successfully captured the numerical instability at $\pi$ radians, although this took around ten times more epochs than other trajectories. 


\subsection{Hénon--Heiles system}\label{henon}
\subsubsection{Statement of the equation of motion and cost function}
The Hénon--Heiles system \cite{henon} describes the non-linear motion of a star around its galactic centre. The motion can be viewed as a pair of harmonic oscillators with a perturbation (associated with a galactic potential) which couples the oscillators together. The motion is chaotic and non-dissipative. This system can, again, be described in terms of a Lagrangian of the form $L=T-V$, where the kinetic energy is
\begin{equation}
  T(\dot{x}(t), \dot{y}(t)) = \frac{1}{2}m\left[\dot{x}^2(t) + \dot{y}^2(t)\right],
\end{equation}
and the potential is
\begin{equation}
  V(x(t),y(t)) = \frac{1}{2}m\omega^2\left[x^2(t)+y^2(t)\right]+k\left[ x^2(t) y(t) - \frac{y^3(t)}{3} \right].
\end{equation}
Ensuring a dimensionless system of equations by setting $m=\omega=k=1$, and by solving the Euler--Lagrange equations, we obtain the system of coupled equations of motion for each coordinate
\begin{equation}
  \begin{aligned}
  \ddot{x}(t) + x(t) - 2 x(t)y(t) &= 0,\\
  \ddot{y}(t) + y(t) + y^2(t) - x^2(t) &=0.
  \end{aligned}
\end{equation}
The system is time-independent, and so the total energy $E=T+V$ is a conserved quantity.

The equations of motion and initial conditions lead to the cost function
\begin{equation}
  \begin{aligned}
    \mathcal{C} &= \| \mathbf{\ddot{x}}(\mathbf{t}) + \mathbf{x}(\mathbf{t}) -2\mathbf{x}(\mathbf{t})\mathbf{y}(\mathbf{t}) \|^2 \\
    &\quad+ \| \mathbf{\ddot{y}}(\mathbf{t}) + \mathbf{y}(\mathbf{t}) + \mathbf{y}^2(\mathbf{t}) - \mathbf{x}^2(\mathbf{t}) \|^2 \\
    &\quad+ (x_0-s_{x,0})^2 + (y_0-s_{y,0})^2\\
    &\quad+ (\dot{x}_0-v_{x,0})^2 + (\dot{y}_0-v_{y,0})^2.
  \end{aligned}
  \label{eq:hh_cost}
\end{equation}

\subsubsection{Coupled neural networks}
We have two dynamical degrees of freedom, which we handle by employing two neural networks --- one for each coordinate --- with separate parameterisations that we simultaneously optimise. The cost function is shared between the two networks, although note that there are two separate backwards propagations of errors needed to update the parameters of each network. The gradients computed during both backward passes are based on the cost function in Eq.~(\ref{eq:hh_cost}), which includes terms for both coordinates. Because these coordinates are coupled in the equations of motion, the gradient of the cost function with respect to one set of parameters will naturally include contributions from both coordinates. The presence of coupling in the cost function means that we must ensure that the updates to the parameters for one network reflect the influence of the other network. The first backward propagation computes the gradients in the network representing $x$; the gradient information is retained after the errors associated with the first coordinate are computed. The second backward propagation computes the gradients in the network representing $y$; all gradient information is then discarded since all backwards passes through all coordinates are complete.

Once we calculate the gradients, the optimiser takes a step to update the parameters in the direction that minimises the cost function. The process of zeroing (or resetting) the gradients, calculating new gradients, and then stepping the optimiser is repeated many times in a training loop until the networks converge to a solution that minimises the cost function. In principle, the coupled optimiser could include as many optimisers as computer memory allows. We refer the reader to our code \cite{griffiths} for further implementation details.

\begin{figure}[h!]
  \includegraphics[width=\linewidth]{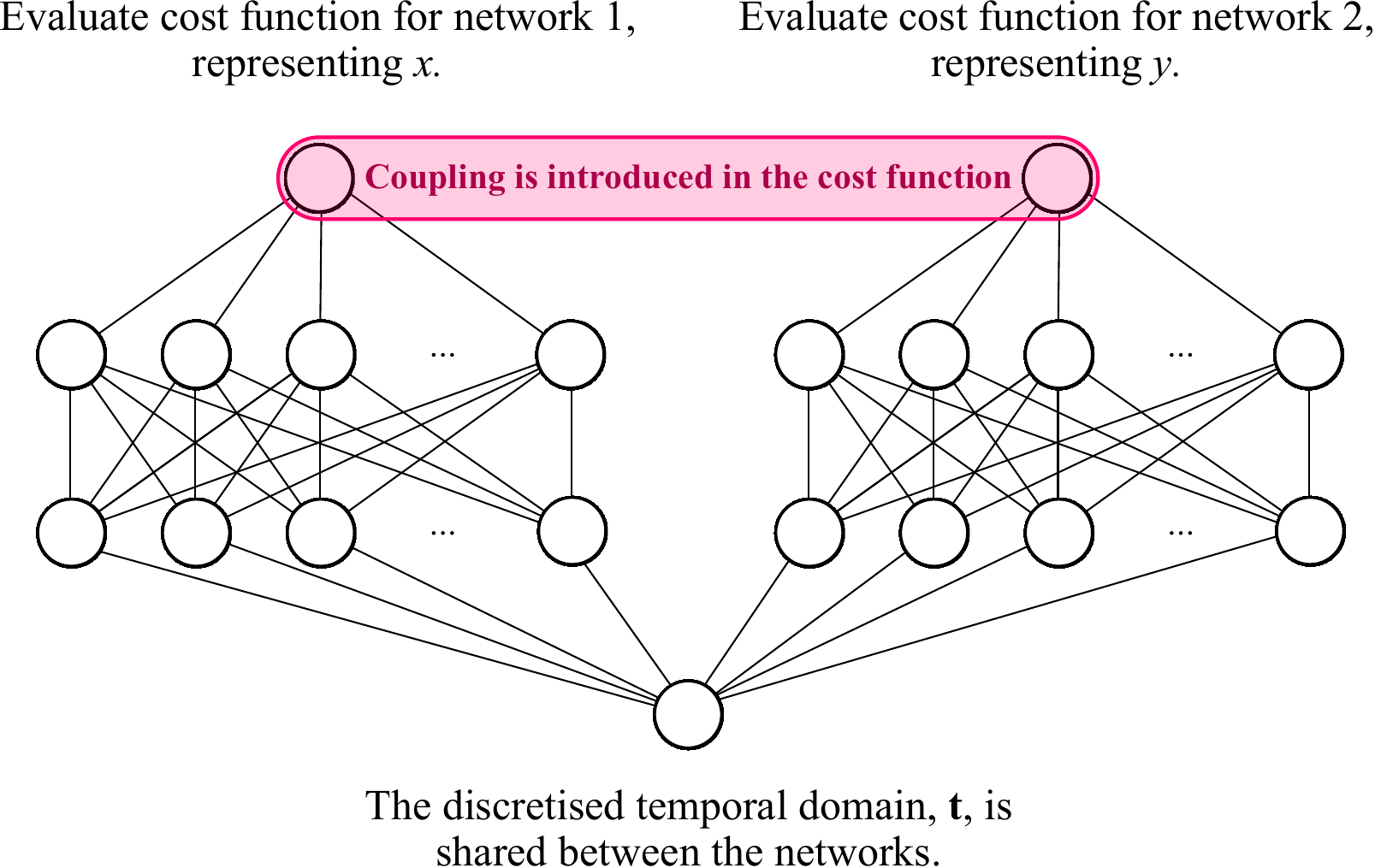}
  \caption[Coupled neural network]{A coupled neural network for the Hénon--Heiles system. It propagates independent parameters for each generalised coordinate (given a common temporal domain) and couples only in the evaluation of the cost functions.}
  \label{fig:coupled_nn}
\end{figure}

Figure \ref{fig:coupled_nn} shows a schematic for the coupled neural network for the Hénon--Heiles system.

\subsubsection{Results and discussion}

\begin{figure}[ht!]
  \subfloat[\label{fig:henon_1}]{%
  \includegraphics[width=0.49\linewidth]{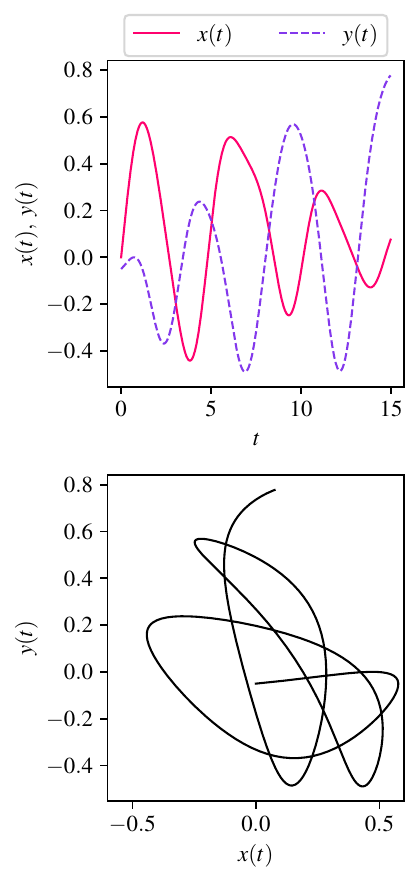}}
  \subfloat[\label{fig:henon_2}]{%
  \includegraphics[width=0.5\linewidth]{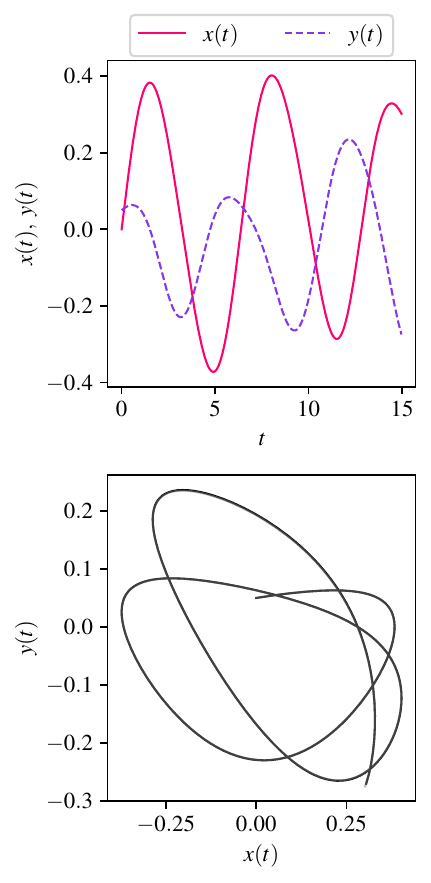}}
  \caption[Neural solutions to the Hénon--Heiles problen.]{Two Hénon--Heiles trajectories obtained by machine learning with different energies: (a) corresponds to $E=1/6$ (the chaotic trajectory) and (b) corresponds to $E=1/12$ (the quasi-periodic trajectory). We show comparative metrics for the same trajectories obtained by Runge--Kutta methods in Fig. (\ref{fig:2_loss_metrics_1_6}) and Fig. (\ref{fig:2_loss_metrics_1_12}). \textbf{Top}: the time evolution of the individual coordinates. \textbf{Bottom}: phase profiles, showing $(x,y)$ pairs as parametric functions of $t$.}
\end{figure}

\begin{figure}
  \includegraphics[width=\linewidth]{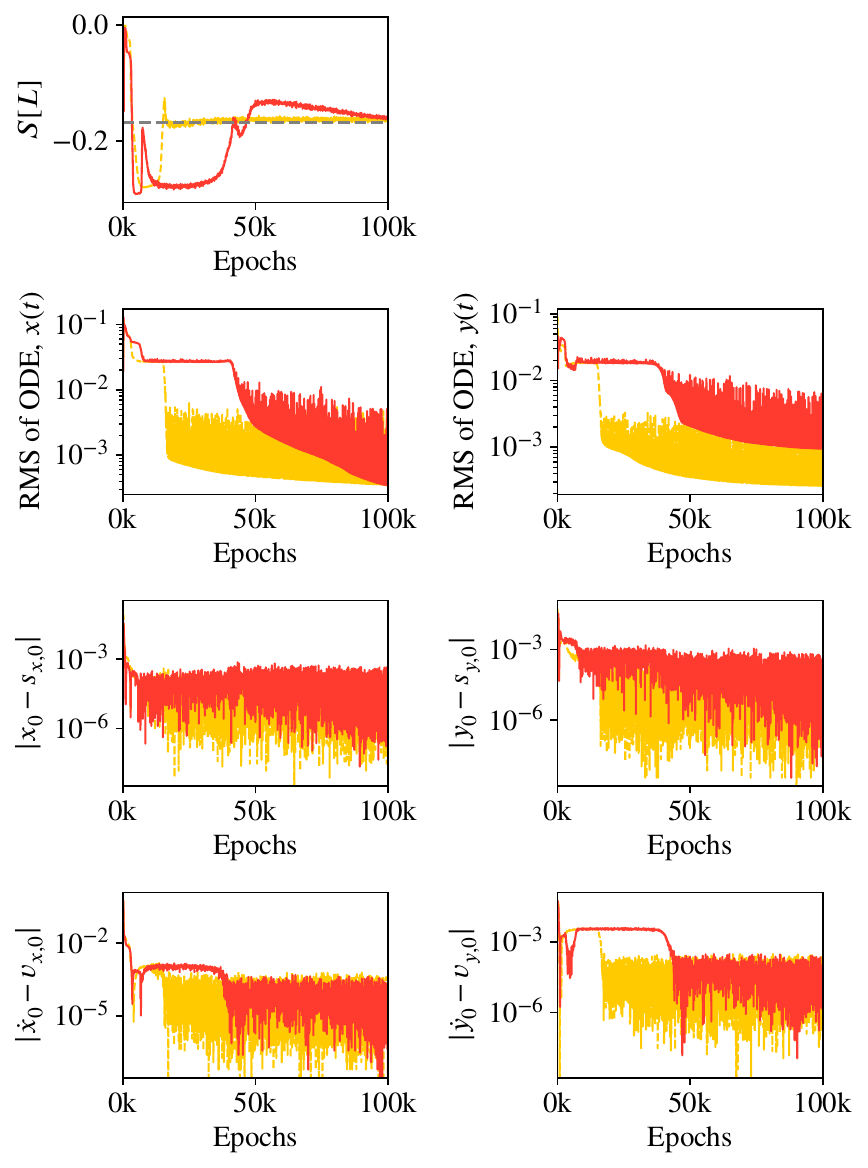}
  \caption[Cost metrics for the Hénon--Heiles problem]{\emph{Hénon--Heiles system for energy $E=1/6$}. \textbf{Row 1:} the evolution of the action as training progresses. The dashed, grey horizontal lines indicate the action of the system obtained by using a Runge--Kutta 4th-order integration scheme. The dashed, yellow trajectories are associated with $N_T=512$ timesteps. The solid, orange trajectories are associated with $N_T=256$ timesteps. As the training evolves, an extremum of the action is also found, taking similar values to the Runge--Kutta integration schemes. \textbf{Row 2:} the root mean square of the residual of the differential equations for the respective coordinates. \textbf{Row 3:} the residuals in the initial positions for both coordinates. \textbf{Row 4:} the residuals in the initial velocities for both coordinates. The initial conditions are easier to match since the solution begins to diverge as the time coordinate increases.}
  \label{fig:2_loss_metrics_1_6}
\end{figure}

\begin{figure}
  \includegraphics[width=\linewidth]{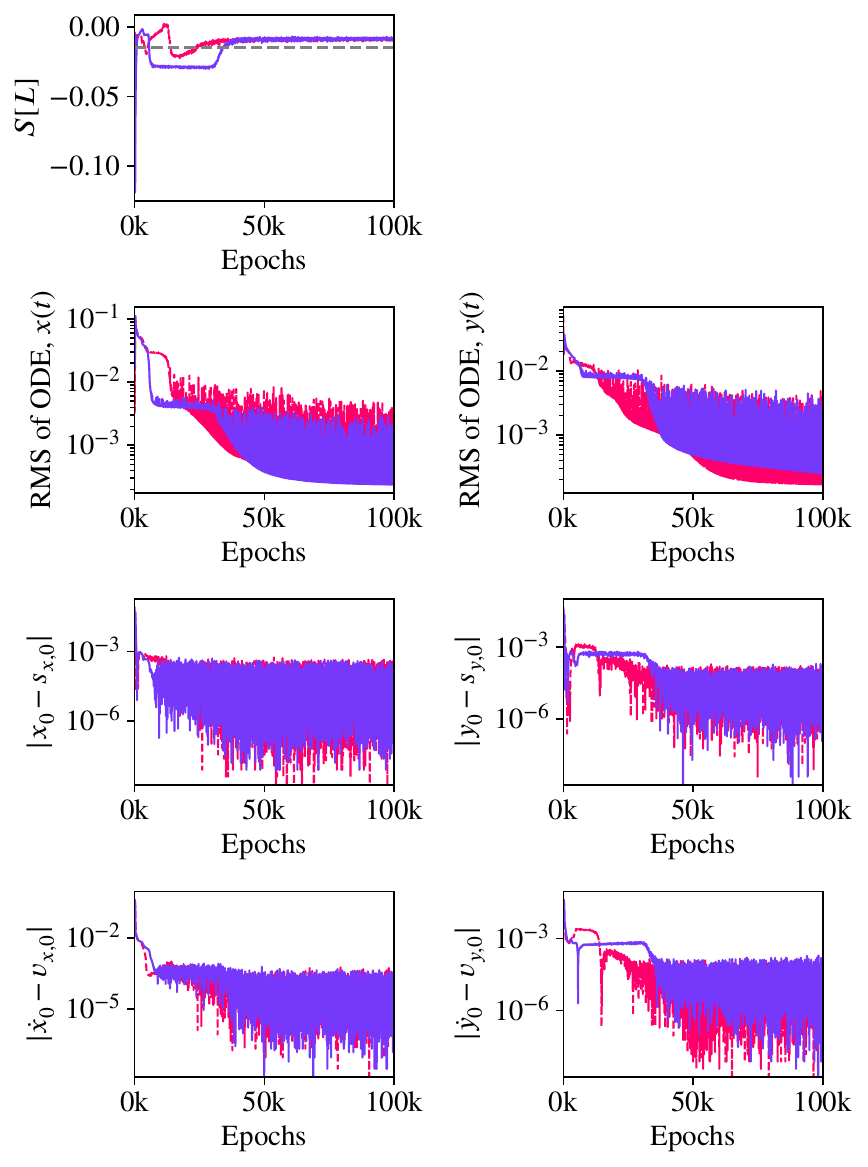}
  \caption[Cost metrics for the Hénon--Heiles problem]{\emph{Hénon--Heiles system for energy $E=1/12$}. \textbf{Row 1:} the evolution of the action as training progresses. The dashed, grey horizontal lines indicate the action of the system obtained by using a Runge--Kutta 4th-order integration scheme. The dashed, pink trajectories are associated with $N_T=512$ timesteps. The solid, purple trajectories are associated with $N_T=256$ timesteps. As the training evolves, an extremum of the action is also found, taking similar values to the Runge--Kutta integration schemes. \textbf{Row 2:} the root mean square of the residual of the differential equations for the respective coordinates. \textbf{Row 3:} the residuals in the initial positions for both coordinates. \textbf{Row 4:} the residuals in the initial velocities for both coordinates. The initial conditions are easier to match since the solution begins to diverge as the time coordinate increases.}
  \label{fig:2_loss_metrics_1_12}
\end{figure}

We will only consider confined (closed) trajectories. Hénon and Heiles give an upper limit for the energy to produce a confined trajectory as $E=1/6$. This trajectory is entirely chaotic. We also study the confined trajectory with energy $E=1/12$, which is quasi-periodic. We choose always to set the initial position of the first coordinate such that $s_{x,0}=0$. For the chaotic trajectory, where $E=1/6$, we set the initial velocity for the $x$-coordinate to $v_{x,0}=0.10$, and the initial position and velocity for the $y$-coordinate to $s_{y,0}=0.57$ and $v_{y,0}=0.057$. For the quasi-periodic trajectory, where $E=1/12$, we set the velocity for the $x$-coordinate to $v_{x,0}=0.05$, and the initial position and velocity for the $y$-coordinate to $s_{y,0}=0.40$ and $v_{y,0}=0.041$.\footnote{The initial velocity of the first coordinate is constrained by $\dot{x}(0)=\sqrt{2E-v_{y,0}^2-s_{y,0}^2+2s_{y,0}^3/3}$, where we have set either $E=1/6$ or $E=1/12$. Given these constraints, the initial position and initial velocity for the second coordinate can be arbitrarily chosen, providing that $\dot{x}(0)$ remains real-valued. To ensure an appropriate choice of the initial conditions for the second coordinate, we find $y(0)$ by using root finding algorithms on the constraint $V<E$, and setting the result as 10\% of this value. We find $\dot{y}(0)$ by evaluating $T<E-V$, and setting the result as 10\% of this value.}



Figures \ref{fig:henon_1} and \ref{fig:henon_2} show these two Hénon--Heiles trajectories. We used the SechLU activation function, given in Eq. (\ref{eq:1_sechlu}). GELU, given in Eq. (\ref{eq:1_gelu}), was not conducive to learning and all trajectories failed to obtain correct dynamics; we are unsure why this is the case, but it is our empirical observation that SechLU is the most reliable of the probabilistic activation functions that we have tested. Figures \ref{fig:2_loss_metrics_1_6} and \ref{fig:2_loss_metrics_1_12} show the individual cost metrics for a Hénon--Heiles system with energies $E=1/6$ and $E=1/12$, respectively.

The neural solutions successfully minimise the action, $S$, as training progresses (we stress that the action is not an explicit term in the cost function, although it generates the dynamical equations through Hamilton's principle). We approximate the action at every epoch using $S\approx\Delta t \sum_{k=0}^{N_T} L(x_k, \dot{x}_k, y_k, \dot{y}_k)$. We note that the action of the machine learning trajectories does not necessarily increase or decrease monotonically with epoch. We note that one might expect that forming a cost function around the action only and then minimising it would be sufficient, as an almost literal realisation of Hamilton's principle of stationary action (the fact that the action must be extremised rather than strictly minimised notwithstanding). We have not found this to be the case, neither in conjunction with a trajectory's initial values, nor with coordinate values at the path's two endpoints. 
The Euler--Lagrange equations, i.e., the dynamical \textit{consequence\/} of Hamilton's principle, in conjunction with the initial conditions, appear to be necessary, which also have the useful property of the minimising value of the cost function always being equal to zero.
We note that when the root mean square of the differential equation remains constant over many epochs, it corresponds to unchanging values of the action. This is due to the equivalence of the principle of stationary action and the underlying equations of motion.

The trajectories obtained by the Runge--Kutta integration routines may not give the `true' value of the action (especially in systems with chaotic dynamics). However, since multiple methods are approaching similar values of the action, this value could be interpreted as the most stationary. 


We finally note that a finer temporal discretisation appears to improve the learning capabilities of the neural network, especially with the chaotic trajectory, without an increase in network parameters (since the neurons are vector-valued). The stability of probing chaotic dynamics using numerical integration schemes such as the Runge--Kutta methods can be improved (up to a point) by using a finer temporal resolution, which is what we observe in our neural solutions.

\section{Conclusions}
We have presented the neural initial value problem: a machine learning framework which approximates the solutions to a range of physics-based initial value problems in the absence of training data (i.e., an unsupervised learning approach). We have formulated a general cost function that captures the dynamics of a wide range of mechanical systems. Through extensive experimentation, we demonstrated that our approach can successfully model systems with non-linear, coupled, and chaotic behaviours. For the free particle and particle in a gravitational field, our linear neural network provided accurate solutions and served as a linear approximant for problems with non-linear solutions. In the case of the pendulum (for arbitrary angles, i.e., outside the small angle approximation), the deep neural network framework proved effective in capturing non-linear dynamics. For the Hénon--Heiles system, our coupled neural network was successful in capturing the dynamics of the two coordinates, including in chaotic regimes. We have also introduced and evaluated several machine learning techniques to enhance our framework: probabilistic activation functions, which appear to be necessary to learn the solutions of initial value problems (they may be interpreted as a general class of model averaging functions), and the coupled optimisation routines allow for the simultaneous optimisation of many neural networks representing individual generalised coordinates.

Whilst not a direct replacement for standard numerical integration routines such as the Runge--Kutta methods, we have demonstrated that neural initial value problems are a complementary approach which may be used to verify the validity of other numerical methods, for example in highly chaotic regimes. Whilst not presented in the current work, we have also found that coupled neural networks --- alongside the neural initial value problem framework --- can solve complex-valued initial value problems by treating the underlying differential equations as a system of ordinary differential equations for the real and imaginary parts (i.e., considering the complex domain to be an ordered pair of real numbers). The methods presented in this paper could be augmented with the presence of training data, where the probabilistic activation functions could provide greater accuracy and/or generalisation capabilities for a broader range of dynamical systems. 

Future work may also be directed towards understanding the non-monotonicity of the action in the optimisation procedure: for example whether the neural network settles temporarily on a different solution (which one might consider to be an ODE equivalent of `hallucination' phenomena observed in AI solutions more generally) obeying Hamilton's principle, possibly one which satisfies the contemporaneous approximation for the initial conditions. Another question of principle is whether a strict action minimisation procedure (choosing two endpoints and then trying to find the minimal action between the points) can be implemented using the framework in this paper.   
\section{Acknowledgements}
We thank the EPSRC, grant EP/T015241/1, for funding this project.

\section{Code availability}
The code for this project is available on GitHub \cite{griffiths}.

\appendix
\section{Output of a scalar-valued neuron} \label{ch1:deep_learning} \label{sec:deep_learning}
In the main text we consider vector neurons exclusively. Here we briefly present the notation for scalar neurons, with the intention of making the parallels between scalar and vector neurons readily apparent.

\begin{figure}[h!]
  \centering
  \includegraphics[width=\linewidth]{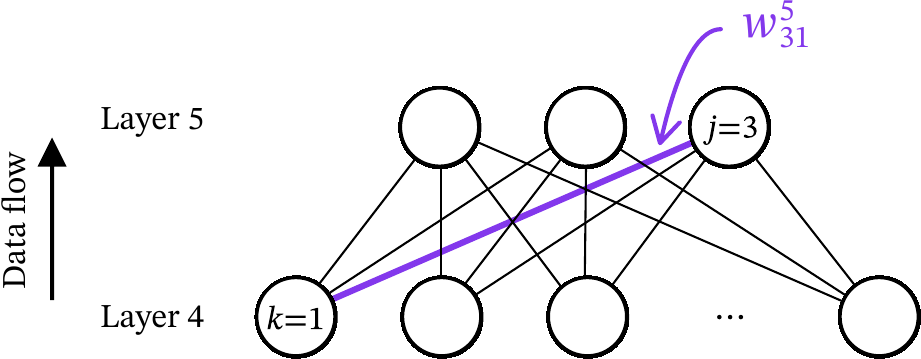}
  \caption[Demonstration of the notation for the weights, $w_{jk}^\ell$, in a feedforward neural network]{Two layers of a feedforward neural network are shown. A connection between the first neuron of the fourth layer and the third neuron of the fifth layer is highlighted. This connection has a weight $w_{jk}^\ell=w_{31}^5$.}
  \label{fig:understanding_weights}
\end{figure}

Let the $k$th neuron from the $(\ell-1)$th layer have an output value of $y_k^{\ell-1}$ and let the $j$th neuron from the $\ell$th layer have an output value of $y_j^\ell$. Let $w_{jk}^\ell$ be the weight from the $k$th neuron in the $(\ell-1)$th layer to the $j$th neuron in the $\ell$th layer. We assume that the neural network is fully connected (all neurons in one layer are connected to all neurons in the subsequent layer) and feedforward (there are no loops or recurrent connections). Figure \ref{fig:understanding_weights} shows this notation between two layers of a neural network. \index{neural network! feedforward}

Using this notation, the output, $y_j^\ell$, of the $j$th neuron in the $\ell$th layer then evaluates as
\begin{subequations}
\begin{align}
  z_j^\ell &= \sum_{k=1}^{n_{\ell-1}} w_{jk}^\ell y_k^{\ell-1} +b_j^\ell,
  \label{ch1:preactivation}
  \\
  y_j^\ell &= \mathcal{A}\left(z_j^\ell\right),
  \label{ch1:postactivation}
\end{align}
\end{subequations}
where Eq. (\ref{ch1:preactivation}) is akin to a linear regression and Eq. (\ref{ch1:postactivation}) is referred to as the activation or output of a neuron in the deep neural network.

Activation functions are introduced between layers to invoke the \emph{universal approximation theorem} \cite{hornik, pinkus, leshno} --- a statement that, with the addition of nonlinear activation functions, a possibly infinitely wide neural network trained for possibly infinite time can represent any continuous function\footnote{The universal approximation theorems do not tell us \emph{how} to find the optimal weights and biases, only that they exist.}. In general, it is not obvious \emph{a priori\/} which activation function to choose; an activation function suitable for one problem is not necessarily suitable for a different problem, as a consequence of the \emph{No Free Lunch\/} theorem \cite{wolpert}.

\section{Probabilistic activation functions}
\subsection{Model averaging and dropout} \label{model_avg}
Neural networks exhibit run-to-run variation --- different initialisations of the network parameters and stochastic descent through the (possibly vast) parameter space mean that each neural network will return a different output compared to another instance of a neural network with the same architecture, cost function, inputs and hyperparameters. Different instances of the neural network explore different local minima, which in turn produce different sets of optimal parameters. It may be that some runs fail to learn the mapping between input and output, whilst others learn a more accurate mapping. It has been noted extensively \cite{breiman, dietterich, hansen, krogh, opitz} that performing an average (arithmetic or geometric mean) of many neural networks produces a more accurate model than any individual neural network.

\begin{figure}[h!]
  \centering
  \subfloat[\label{subfig:a}]{%
  \includegraphics[width=0.4\columnwidth]{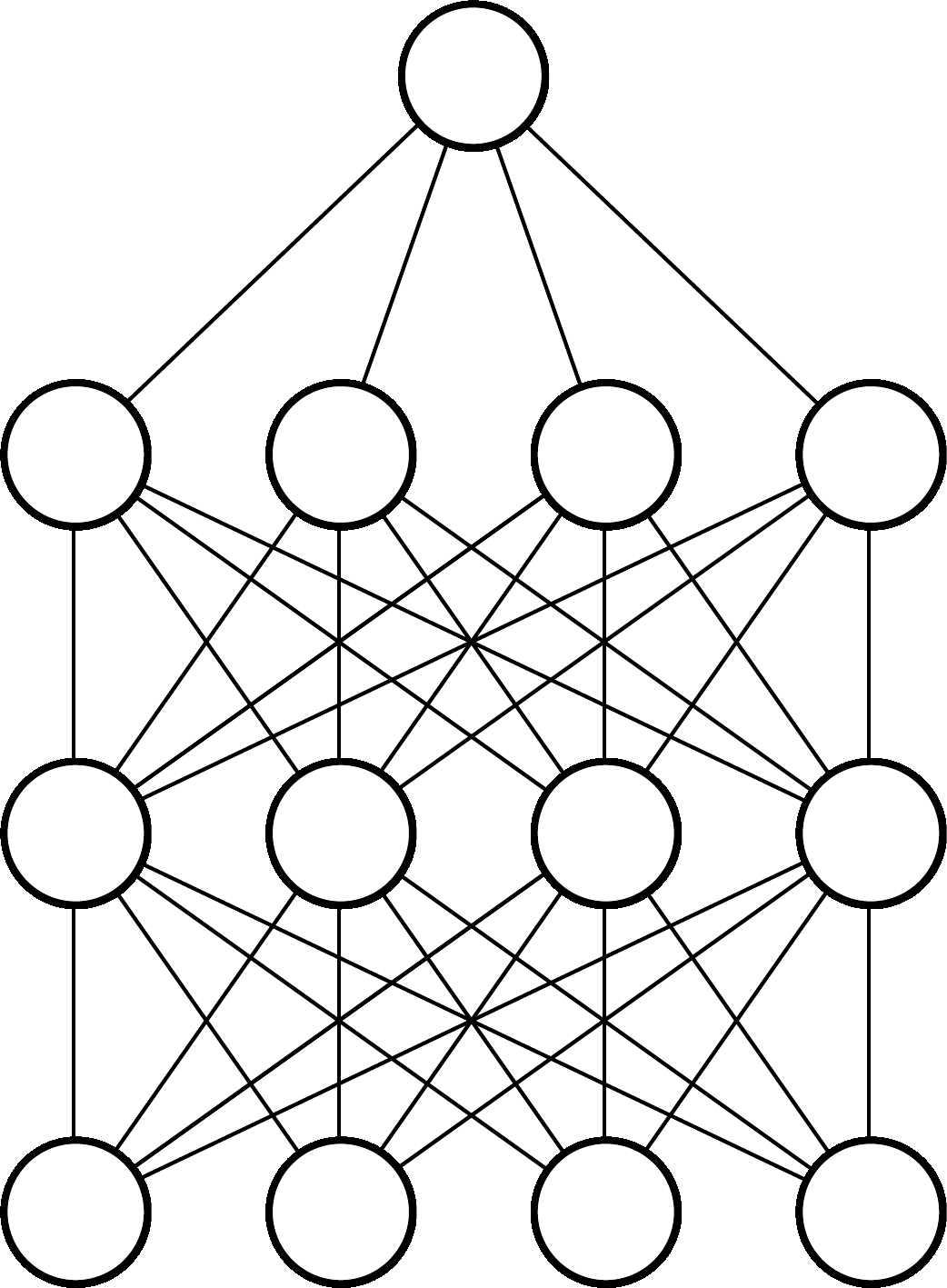}%
  }\quad\quad\quad
  \subfloat[\label{subfig:b}]{%
  \includegraphics[width=0.4\columnwidth]{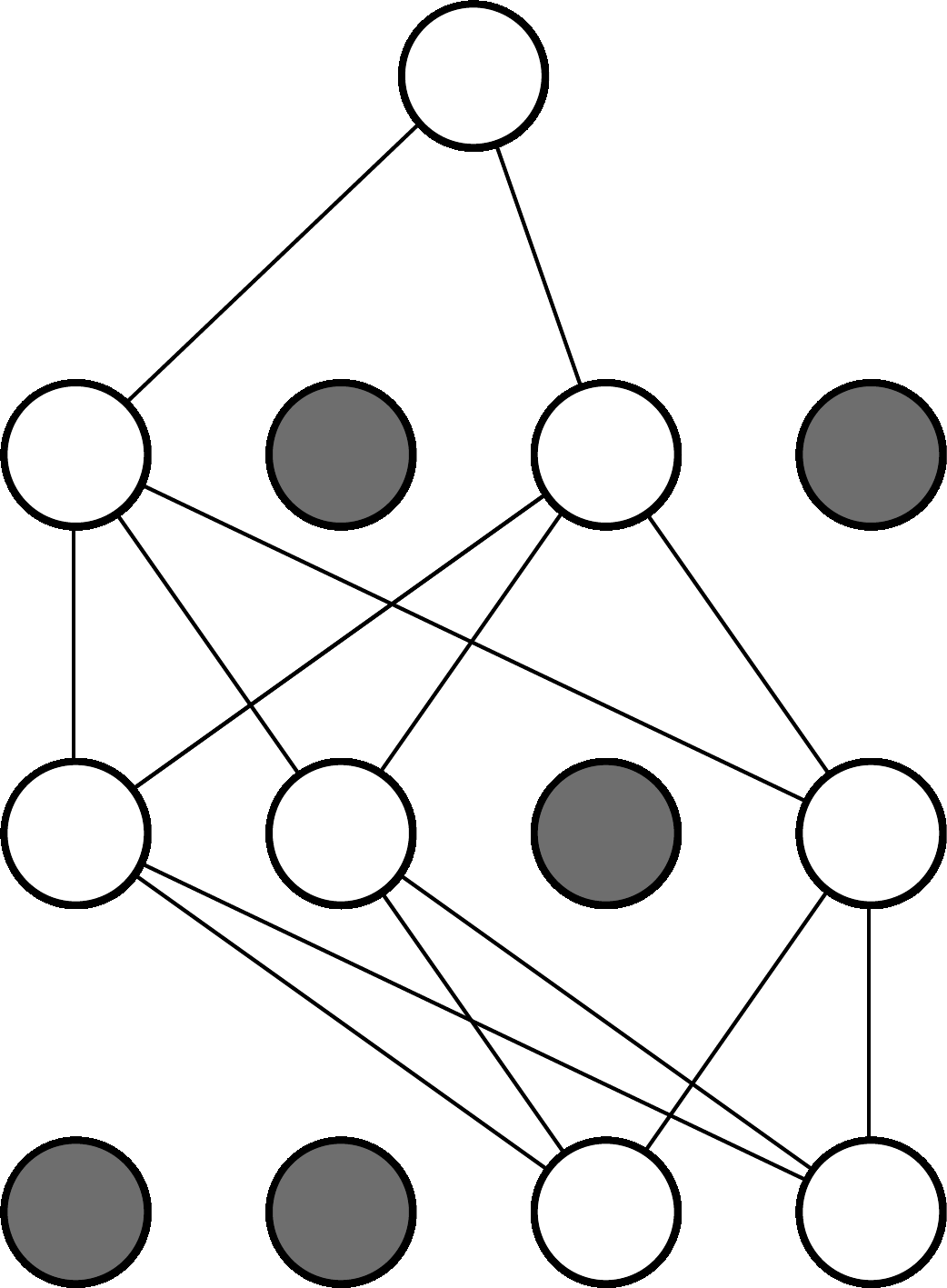}%
  }
  \caption[Dropout (with and without)]{(a) A fully connected, feedforward neural network without dropped neurons and (b) the same neural network with dropped neurons (dark grey) and the remaining connections. The neurons which are dropped are stochastically chosen at each epoch.}
  \label{fig:understanding_dropout}
\end{figure}

Model averaging may be heuristically achieved using an algorithm called dropout. Consider the output of the $j$th neuron in the $\ell$th layer, given in Eqs.~(\ref{ch1:preactivation}) and (\ref{ch1:postactivation}). Dropout, introduced by Srivastava \emph{et al.} \cite{srivastava}, removes neurons with probability $p$ and preserves neurons with probability $1-p$ at the start of each epoch. We can represent the effect of dropout through a set of Bernoulli trial outcomes $r_k^{\ell-1}$ ($=0$ with probability $p$ and otherwise $=1$). We then define $\tilde{y}_k^{\ell-1} = r_k^{\ell-1} y_k^{\ell-1}$, such that, instead of Eq.~(\ref{ch1:preactivation}), we have
\begin{equation}
  z_j^{\ell} = \sum_{k=1}^{n_{\ell-1}} w_{jk}^\ell \tilde{y}_k^{\ell-1} +b_j^\ell,
  \label{eq:dropout}
\end{equation}
with  $y_j^\ell = \mathcal{A}(z_j^\ell)$, as in  Eq.~(\ref{ch1:postactivation}). 
Figure \ref{fig:understanding_dropout} shows a feedforward neural network, with and without dropout. Configurations with dropout are called subnetworks, as they are a subset of the larger network. 
Neurons in all layers except the output layer can be subject todropout. It is in principle possible that there are no connections between the input and output layer, but this becomes vanishingly unlikely for larger networks.

Dropout prevents overfitting (referred to as regularisation), and also approximately averages the output of multiple neural networks. Dropout is carried out at the start of each epoch. 
For a network with $N$ neurons in the hidden layers, there are $2^N$ possible configurations of the neurons in the network. Dropout is thus akin to an ensemble of $2^N$ instances of a neural network, and can be described as an average of many models produced by the subnetworks.

A process somewhat analogous to dropout can be carried out via a class of activation functions we call probabilistic activation functions, extending work by Hendrycks and Gimpel \cite{hendrycks}. 
It is our observation that probabilistic activation functions are required for learning solutions of initial value problems.

\subsection{Connection to probabilistic activation functions}\label{sec:1_probabilistic_act} \index{activation functions!probabilistic}
We define the \emph{probabilistic activation function}
\begin{equation}
  \mathcal{A}(x) = x\phi(x) =  x\int_{-\infty}^x \mathrm{d}x'\,p(x'),
\end{equation}  
where $p(x)$ is a probability distribution supported on the domain $x\in(-\infty,\infty)$ and the integral defines its cumulative distribution function $\phi(x)$  \footnote{Probability distributions with support only on the domain $x\in(0,\infty)$ are not conducive to learning, possibly due to the lack of symmetry. It is our empirical observation that some distributions which are not perfectly symmetric about the origin --- such as that used in Mish (see Eq. (\ref{eq:1_mish})) --- are still conducive to learning.}.

Although not smooth or differentiable, it is insightful to consider the Dirac delta distribution, $p(x)=\delta(x)$. The associated cumulative distribution function is the Heaviside step function,
\begin{equation}
  \phi(x) = \int^x_{-\infty} \mathrm{d}x'\, \delta(x') = \Theta(x),
\end{equation}
which leads to the standard (and also non-differentiable) ReLU activation function\index{activation functions!Rectifying error linear unit (ReLU)}
\begin{equation}
  \mathrm{ReLU}(x)=x\mathop{\Theta(x)}=\max(0,x).
  \label{eq:1_relu}
\end{equation}
The effect is therefore comparable to a form of dropout; if $x$ is viewed as a random variable, then neurons are either preserved or dropped, with probability $0.5$.
More generally, the cumulative distribution function $\phi(x)$ for a continuous, differentiable distribution function $p(x)$ will interpolate smoothly between being close to $0$ over much of the initial part of its range, and close to $1$ over the latter part of its range. Note also that if $\phi(x)$ is the probability for the outcome of a Bernoulli trial to be $=1$, it is also the expectation value for the Bernoulli trial. Hence, a probabilistic activation function $\mathcal{A}(x)=x \mathop{\phi(x)}$ 
scales all neuron outputs by the expectation value of a Bernoulli trial, effectively providing something resembling a smooth implementation of dropout and a form of model averaging.

\begin{figure*}
  \centering
  \includegraphics[width=\linewidth]{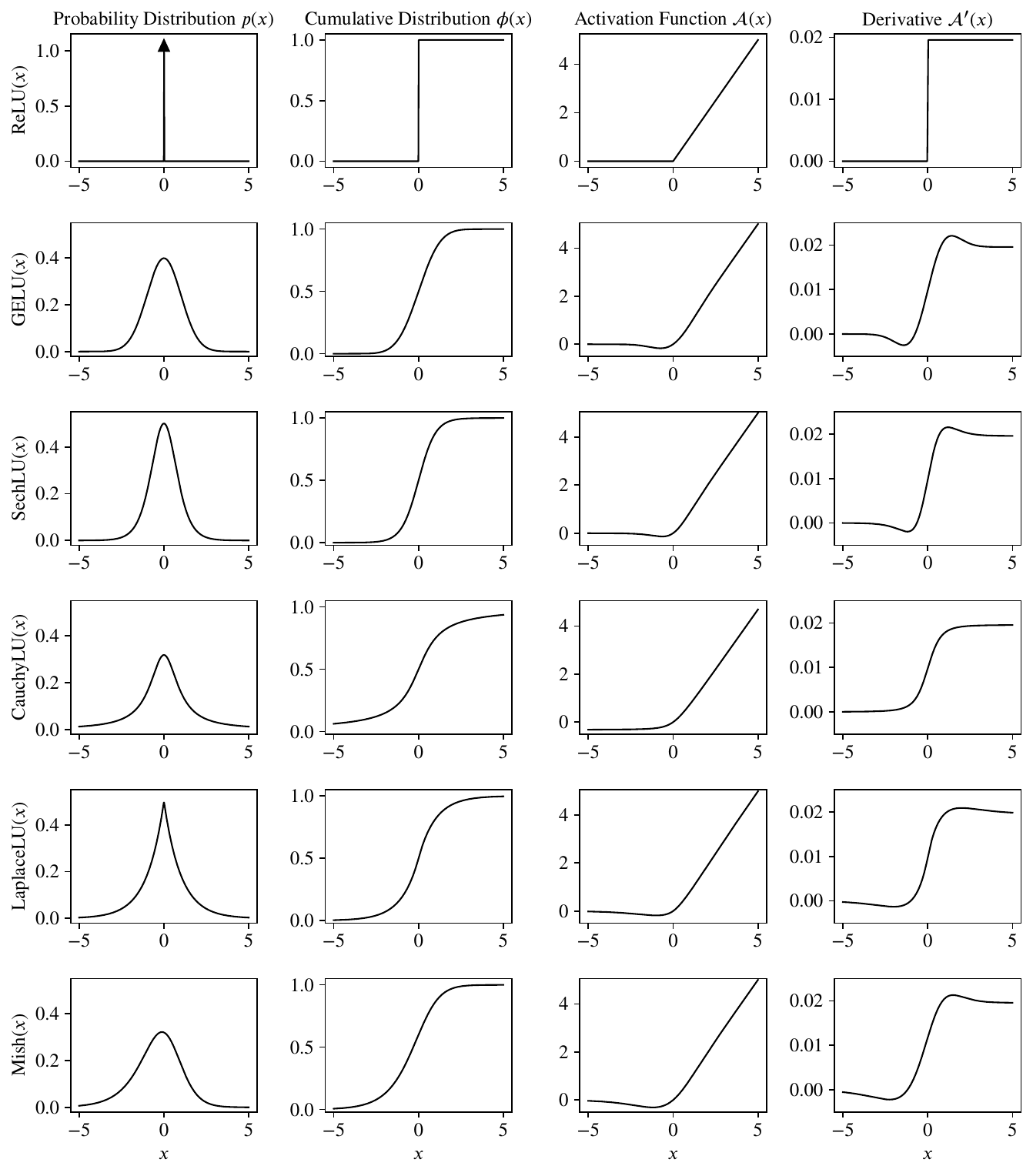}
  \caption[Plots of probabilistic activation functions, their probability distributions, cumulative distributions, and derivatives.]{A selection of probabilistic activation functions are shown: the ReLU (Eq. (\ref{eq:1_relu})), GELU (Eq. (\ref{eq:1_gelu})), SechLU (Eq. (\ref{eq:1_sechlu})), CauchyLU (Eq. (\ref{eq:1_cauchylu})), LaplaceLU (Eq. (\ref{eq:1_laplacelu})) and Mish (Eq. (\ref{eq:1_mish})). All figures use either the variance or scaling parameter $\rho=1$ (even if this is not the most optimal value). The first column is the probability distribution $p(x)$. The second column is the cumulative distribution $\phi(x)=\int_{-\infty}^x \mathrm{d}t p(t)$. The third column is the probabilistic activation function $\mathcal{A}(x)=x\phi(x)$. The fourth column is the derivative of the activation function $f'(x)$. The functions, with the exception of ReLU, are plotted on common axes to see clearly the relative magnitude and curvature of the functions.}
  \label{fig:1_activations}
\end{figure*}

In this way one can derive the Gaussian error linear unit (GELU) activation function \cite{hendrycks} \index{activation functions!Gaussian error linear unit (GELU)}
\begin{subequations}
  \begin{align}
  p(x) &= \frac{1}{\rho\sqrt{2\pi}}\mathrm{exp}\left(-\frac{x^2}{2\rho^2}\right), \\
  \mathcal{A}(x) &= \mathrm{GELU}(x) =  \frac{x}{2}\left[1+\mathrm{erf}\,\left(\frac{x}{\rho\sqrt{2}}\right)\right].
  \label{eq:1_gelu}
  \end{align}
\end{subequations}
An activation function 
we have found in this work to be very useful
we call SechLU: 
\index{activation functions!hyperbolic secant linear unit (SechLU)}
\begin{subequations}
\begin{align}
  p(x) &= \frac{1}{2\rho}\mathrm{sech}^2\left(\frac{x}{\rho}\right), \\
  \mathcal{A}(x) &= \mathrm{SechLU}(x) = \frac{x}{1+\mathrm{exp}(-2x/\rho)}.
  \label{eq:1_sechlu}
\end{align}
\end{subequations}
Starting with a Cauchy (Lorentzian) distribution, we
can 
similarly
obtain an activation function  we call CauchyLU:
\begin{subequations}
  \begin{align}
  p(x) &= \frac{1}{\pi\rho} \left( 1+\frac{x^2}{\rho^2} \right)^{-1},  \\
  \mathcal{A}(x) &= \mathrm{CauchyLU}(x) = \frac{x}{2}\left[1 + \frac{2}{\pi}\arctan\left(\frac{x}{\rho}\right)\right].
  \label{eq:1_cauchylu}
  \end{align}
\end{subequations} \index{activation functions!Cauchy linear unit (CauchyLU)}
Noting it is not everywhere differentiable, we
can obtain an activation function which we call LaplaceLU: 
\begin{subequations}
\begin{align}
  p(x) &= \frac{1}{2\rho} \mathrm{exp}\left( -\frac{|x|}{\rho} \right) \\
\mathcal{A}(x) &= \mathrm{LaplaceLU}(x) = \frac{x}{2}\left\{ 1+ \operatorname{sgn}(x)\left[1-\exp \left(-\frac{|x|}{\rho}\right)\right]\right\}.
  \label{eq:1_laplacelu}
\end{align}  
\end{subequations}\index{activation functions!Laplace linear unit (LaplaceLU)}
Finally, the Mish activation function, introduced by Misra in 2019 \cite{misra}, can also be derived from a (somewhat unusual) probability distribution. 
\begin{subequations}
  \begin{align}
  p(x) &= \frac{1}{\rho}\mathop{\mathrm{sech}^2\left( \ln\left|1+\mathrm{e}^{x/\rho}\right| \right)} \frac{\mathrm{e}^{x/\rho}}{1+\mathrm{e}^{x/\rho}},  \\
\mathcal{A}(x) &= \mathrm{Mish}(x) = x \mathop{\tanh\left( \ln\left|1+\mathrm{e}^{x/\rho}\right| \right)}.
  \label{eq:1_mish}
  \end{align}
\end{subequations} \index{activation functions!Mish}
The probability distributions, cumulative distributions, activation functions and their derivatives are shown 
in Fig. \ref{fig:1_activations}.


In all cases, the 
width parameter $\rho$ may be learnt as an additional hyperparameter by following the same initialisation and optimisation routines as for the weights and biases. We observe faster and more reliable convergence to appropriate physical solutions if we treat $\rho$ as a free parameter.

\section{Weight and bias initialisation} \index{initialisation} \label{ch1:initialisation}

The weights and the biases in the neural network must be initialised before any training or learning can take place, where it is important to note that this is not a ``best guess'' at the solution. The values for this parameter initialisation should be neither too large (to avoid diverging values of the gradients) nor too small (to avoid slow training).

In what follows we outline an initialisation scheme general to any differentiable activation function based on that described by Glorot and Bengio \cite{xavier}, which assumed the consistent use throughout the network of the hyperbolic tangent. Without loss of generality we may ignore the bias terms to simplify the mathematics (the biases are, in any case, drawn from the same distribution as the weights). We assume the weights distribution has mean $=0$ and variance $=\sigma^2$, and that the inputs $y_k^{\ell-1}$ are distributed with mean $=0$ and variance $=\rho^2$ (in general $\sigma^2 \neq \rho^2$).
The parameters are initialised with zero mean to avoid erratic and undesirable zig-zagging through the parameter space during optimisation, which may cause overshooting of the minima. The requirement for constant variance, i.e., $\mathrm{Var}(y_k^{\ell-1}) \approx \mathrm{Var}(y_j^{\ell})$, prevents the layers' output values from exponentially vanishing or exploding as one moves from bottom to top through the network, which would make the network difficult or impossible to train.

Early on in the training, the values of $z_j^\ell$ can be considered to be small \cite{xavier}, such that the action of a differentiable activation function can be approximated by $y_j^\ell = \mathcal{A}(z_j^\ell) \approx \mathcal{A}(0) + \mathcal{A}'(0) z_j^\ell$. We can therefore consistently approximate the expectation value by
\begin{equation}
  \mathbb{E}(y_j^\ell)
  \approx \mathcal{A}(0)+\mathop{\mathcal{A}'(0)}\mathop{\mathbb{E}(z_j^\ell)},
\end{equation}
and the variance $\mathrm{Var}(y_j^\ell)=\mathbb{E}[(y_j^\ell)^{2}]-[\mathbb{E}(y_j^\ell)]^{2}$ by
\begin{equation}
 \mathrm{Var}(y_j^\ell)
  \approx \mathcal{A}'(0)^{2}
  \mathop{\mathrm{Var}}(z_j^\ell).
\end{equation}
Hence, assuming $w_{jk}^\ell$ and $y_k^{\ell-1}$ to be independent random variables, $\mathbb{E}(w_{jk}^\ell y_k^{\ell-1})=\mathop{\mathbb{E}}(w_{jk}^\ell) \mathop{\mathbb{E}(y_k^{\ell-1})}$, and
\begin{equation}
  \mathbb{E}(y_j^\ell) \approx \mathcal{A}(0) + \mathcal{A}'(0) \sum_{k=1}^{n_{\ell-1}} \mathop{\mathbb{E}}(w_{jk}^\ell) \mathop{\mathbb{E}(y_k^{\ell-1})}.
\end{equation}
As we have already assumed the distribution from which the weights are drawn, and the distribution of the neuron inputs, both have zero mean, it follows that $\mathbb{E}(y_j^\ell) \approx \mathcal{A}(0)$. Similarly,
\begin{equation}
\begin{split}
\mathrm{Var}(y_j^\ell)
 \approx & \mathcal{A}'(0)^{2}
\sum_{k,k'=1}^{n_{\ell-1}} \mathbb{E} 
(w_{jk}^\ell y_k^{\ell-1} w_{jk'}^\ell y_{k'}^{\ell-1})
\\=&
\mathcal{A}'(0)^{2}
\sum_{k=1}^{n_{\ell-1}} \mathop{\mathbb{E} 
[(w_{jk}^\ell)^{2}] }
\mathop{\mathbb{E} 
[(y_k^{\ell-1})^{2}]},
\end{split}
\end{equation}
as $\mathbb{E}(w_{jk}^\ell y_k^{\ell-1} w_{jk'}^\ell y_{k'}^{\ell-1})
=
\mathop{\mathbb{E}(w_{jk}^\ell)}
\mathop{\mathbb{E}(y_k^{\ell-1})}
\mathop{\mathbb{E}(w_{jk'}^\ell)}
\mathop{\mathbb{E}(y_{k'}^{\ell-1})}
=0
$
if $k\neq k'$ and the variables are therefore all independent. Hence, $\mathrm{Var}(y_j^\ell)
 \approx 
\mathcal{A}'(0)^2 n_{\ell-1} \sigma^2 \rho^2.
 $

For the variance to be the same from layer to layer, i.e., $\mathrm{Var}(y_j^\ell)=\mathrm{Var}(y_k^{\ell-1})= \rho^{2}$, it follows that 
$\mathcal{A}'(0)^2 n_{\ell-1} \sigma^2 =\mathcal{A}'(0)^2 n_\ell \sigma^2 = 1$ (noting that $n_{\ell-1} \neq n_\ell$ in general). We therefore take the average
$\mathcal{A}'(0)^2(n_{\ell-1} + n_\ell) \sigma^2/2 = 1$, from which it follows that 
\begin{equation}
\sigma^2 = \frac{2}{\mathcal{A}'(0)^2 (n_{\ell-1} + n_\ell)}.  \label{eq:generalised_variance}
\end{equation}
Generalised Xavier initialisation uses a uniform (rectangular) distribution over the interval $[-a,a]$, which has mean $=0$ and variance $\sigma^{2}= a^{2}/3$. Equating this with Eq.~(\ref{eq:generalised_variance}) gives the values of the support of the distribution in terms of the number of neurons in each layer
\begin{equation}
  w_{jk}^\ell \sim U\left(-\sqrt{\frac{6}{\mathcal{A}'(0)^2 (n_{\ell-1} + n_\ell)}}, \sqrt{\frac{6}{\mathcal{A}'(0)^2 (n_{\ell-1} + n_\ell)}}\right).
\label{eq:generalised_initialisation}
\end{equation}
For $\tanh(x)$, the scaling factor $\mathcal{A}'(0)=1$. 
For any probabilistic activation function, the scaling factor $\mathcal{A}'(0)=\phi(0)$. If the underlying probability distribution is symmetric, then ${\mathcal{A}'(0)=1/2}$.

Note that the initialisation we employ for linear neural networks in section \ref{linear_net_init} is the Kaiming He scheme, where one samples from an unscaled uniform distribution \cite{he}:
\begin{equation}
  w_{jk}^\ell \sim U\left(-\sqrt{\frac{1}{n_\ell}}, \sqrt{\frac{1}{n_\ell}}\right).
\end{equation}
If $n_\ell=1$, then this reduces to Eq. (\ref{ch3:constant_velocity_initialisation}) in the main text.

\section{Learning algorithms} \label{ch1:learning} \index{learning algorithms}
\subsection{Statement of the problem}
The machine learning problem is to find the set of weights and biases which approximate our solution, and we use our cost function to check if we are converging to the expected solution. 
In all examples in this paper, the weights and biases are adjusted by the adaptive moment estimation or \emph{Adam} optimisation algorithm \cite{kingma}. Adam is the descendent of other optimisation algorithms which have emerged over the last few decades. We will follow Adam's lineage, starting with gradient descent, before arriving at the equations for its implementation. 

\subsection{Gradient descent} \index{learning algorithms!gradient descent} \label{ch1:gradient_descent}
One can extend the gradient descent algorithm introduced in section \ref{gradient_descent} to consider any number of weights and biases. All cost functions are parameterised by their weights and biases $\pmb\theta$. Let us again ignore the possibility of biases in the network and focus on the adjustments required for the weights. For an $L$-layer neural network,
\begin{equation}
  \mathcal{C}(\pmb\theta)\equiv\mathcal{C}(w_{11}^1,\cdots,w_{n_\ell n_{\ell-1}}^\ell,\cdots, w_{n_L n_{L-1}}^L).
\end{equation}
We wish to adjust the weights by some amount $\Delta w_{jk}^\ell$ (and, in general, the biases by some assumed small amount $\Delta b_{j}^\ell$).
These adjustments are described collectively by the vector $\Delta\pmb\theta$. Taylor expanding,
\begin{equation}
  \mathcal{C}(\pmb\theta+\Delta\pmb\theta) \approx \mathcal{C} + (\Delta\pmb\theta)\cdot\nabla_{\pmb\theta}\mathcal{C}(\pmb\theta),
\end{equation}
where by definition
\begin{equation}
  \nabla_{\pmb\theta}\mathcal{C}(\pmb\theta) = \left( \frac{\partial\mathcal{C}(\pmb\theta)}{\partial w_{11}^1}, \cdots, \frac{\partial\mathcal{C}(\pmb\theta)}{\partial w_{n_\ell n_{\ell-1}}^\ell}, \cdots, \frac{\partial\mathcal{C}(\pmb\theta)}{\partial w_{n_L n_{L-1}}^L} \right).
\end{equation}
For $\mathcal{C}\to 0$, we require in general that  $\mathcal{C}(\pmb\theta + \Delta\pmb\theta) < \mathcal{C}(\pmb\theta)$, and that the adjustments point in the opposite direction to the gradient, i.e.,
\begin{equation}
  \Delta\pmb\theta = -\eta\nabla_{\pmb\theta}\mathcal{C}(\pmb\theta),
\end{equation}
where $\eta>0$ is (so far) a fixed global hyperparameter (a parameter which describes a neural network) known as the step size or learning rate. \index{hyperparameter} One, therefore, has the epoch-to-epoch update formulas for the weights and, by corollary, for the biases
\begin{equation}
  \Delta w_{jk}^\ell = - \eta \frac{\partial\mathcal{C}(\pmb\theta)}{\partial w_{jk}^\ell}\quad\text{and}\quad \Delta b_{j}^\ell = -\eta \frac{\partial\mathcal{C}(\pmb\theta)}{\partial b_{j}^\ell}.
  \label{eq:gradient_descent_wb}
\end{equation}

\subsection{Adaptive gradient methods (AdaGrad and RMSProp)} \index{learning algorithms!AdaGrad} \index{learning algorithms!RMSProp}
The cost function may be highly sensitive to certain directions in the parameter space $\pmb\theta$ and insensitive to others. A class of learning algorithms (AdaGrad \cite{duchi}, RMSProp \cite{hinton2} and Adam \cite{kingma}) instead introduce adjustable hyperparameters $\eta_{ij}^\ell$ for every parameter in the neural network to adjust the rate of descent across the parameter space.

AdaGrad adapts the learning rates for each parameter by scaling them by a rolling history of previous gradients $r_{ij}^\ell$,
  \begin{align}
    r_{jk}^\ell &= r_{jk}^\ell + \left[ \frac{\partial\mathcal{C}(\pmb\theta)}{\partial w_{jk}^\ell} \right]^* \left[ \frac{\partial\mathcal{C}(\pmb\theta)}{\partial w_{jk}^\ell} \right],\nonumber \\
    \Delta w_{jk}^\ell &= -\eta\frac{1}{\delta + \sqrt{r_{jk}^\ell}} \left[ \frac{\partial\mathcal{C}(\pmb\theta)}{\partial w_{jk}^\ell} \right],
  \end{align}
where $^*$ denotes the complex conjugate in the case of complex-valued neural networks, $\eta$ is the \emph{global} learning rate and $\delta\sim 10^{-8}$ is a small parameter for numerical stability. The initial values of $r_{jk}^\ell$ are all zero.

RMSProp introduces a decay rate for the accumulated gradients, reducing the dependence on earlier gradients and prioritising the more recent history of the descent through the parameter space (e.g., when RMSProp finds a highly convex part of the parameter space, it will very rapidly converge on the minimum, whereas AdaGrad may never make it into such a region of the parameter space). Let $\rho_1\in[0,1)$ be a decay parameter. The RMSProp algorithm is
  \begin{align}
    r_{jk}^\ell &= \rho_1 r_{jk}^\ell + (1-\rho_1)\left[ \frac{\partial\mathcal{C}(\pmb\theta)}{\partial w_{jk}^\ell} \right]^* \left[ \frac{\partial\mathcal{C}(\pmb\theta)}{\partial w_{jk}^\ell} \right],\nonumber \\
    \Delta w_{jk}^\ell &= -\eta\frac{1}{\delta+\sqrt{r_{jk}^\ell}} \left[ \frac{\partial\mathcal{C}(\pmb\theta)}{\partial w_{jk}^\ell} \right].
  \end{align}

\subsection{Adaptive moment estimation (Adam)} \label{ch1:adam} \index{learning algorithms!Adaptive moment estimation (Adam)}
Adaptive moment estimation (Adam) is an extension of RMSProp and is considered to be the state of the art optimisation algorithm due to its fast and reliable convergence. In addition to the accumulation of the square of the gradients (or square of the modulus of the gradients), Adam also computes an accumulation of the gradients, decayed by some parameter $\rho_2$
  \begin{align}
    r_{jk}^\ell &= \rho_1 r_{jk}^\ell + (1-\rho_1)\left[ \frac{\partial\mathcal{C}(\pmb\theta)}{\partial w_{jk}^\ell} \right]^* \left[ \frac{\partial\mathcal{C}(\pmb\theta)}{\partial w_{jk}^\ell} \right],\nonumber \\
    s_{jk}^\ell &= \rho_2 s_{jk}^\ell + (1-\rho_2) \left[ \frac{\partial\mathcal{C}(\pmb\theta)}{\partial w_{jk}^\ell} \right],\nonumber \\
    \Delta w_{jk}^\ell &= -\eta\frac{s_{jk}^\ell}{\delta+\sqrt{r_{jk}^\ell}},
  \end{align}
where $r_{jk}^\ell$ and $s_{jk}^\ell$ are both initialised to be zero at epoch 0. Since the accumulated gradient and square of the gradients are initially zero, there is a tendency for training to be biased towards zero values, which is undesirable. Adam introduces correction terms to $r_{jk}^\ell$ and $s_{jk}^\ell$ to avoid such bias,
  \begin{align}
    r_{jk}^\ell &= \rho_1 r_{jk}^\ell + (1-\rho_1)\left[ \frac{\partial\mathcal{C}(\pmb\theta)}{\partial w_{jk}^\ell} \right]^* \left[ \frac{\partial\mathcal{C}(\pmb\theta)}{\partial w_{jk}^\ell} \right],\nonumber \\
    s_{jk}^\ell &= \rho_2 s_{jk}^\ell + (1-\rho_2) \left[ \frac{\partial\mathcal{C}(\pmb\theta)}{\partial w_{jk}^\ell} \right],\nonumber \\
    \hat{r}_{jk}^\ell &= \frac{r_{jk}^\ell}{1-\rho_1^\mathcal{E}}\nonumber \\
    \hat{s}_{jk}^\ell &= \frac{s_{jk}^\ell}{1-\rho_2^\mathcal{E}}\nonumber \\
    \Delta w_{jk}^\ell &= -\eta\frac{\hat{s}_{jk}^\ell}{\delta+\sqrt{\hat{r}_{jk}^\ell}},
    \label{eq:adam}
  \end{align}
where lines 3 and 4 are the corrected accumulated gradients (and square of gradients) and the superscript denotes the power of the current epoch number $\mathcal{E}$. As with gradient descent, the biases are updated in a similar way.

\subsection{Backwards propagation of errors with scalar-valued neurons}\label{backprop_scalar}
In all learning algorithms --- from gradient descent to Adam --- we need knowledge of the quantities
\begin{equation}
  \frac{\partial\mathcal{C}}{\partial w_{jk}^\ell}\quad\text{and}\quad \frac{\partial\mathcal{C}}{\partial b_j^\ell},
  \label{eq:backprop_scalar}
\end{equation}
which we refer to as the \emph{sensitivity} of the cost function with respect to a given weight or bias in the network. When we have completed a forward pass through the network, we have a number of outputs which need evaluating against our already defined cost function. The output layer neuron(s) are dependent upon all neurons which come before it. Since a neural network is a composition of many functions, 
we note that the quantities in Eq. (\ref{eq:backprop_scalar}) can be determined from the chain rule and simplified,
\begin{equation}
  \frac{\partial\mathcal{C}}{\partial w_{jk}^\ell} = \frac{\partial\mathcal{C}}{\partial z_j^\ell}\frac{\partial z_j^\ell}{\partial w_{jk}^\ell} = \frac{\partial\mathcal{C}}{\partial z_j^\ell}
  y_k^{\ell-1},
  \label{eq:backprop_wrt_weight}
\end{equation}
and
\begin{equation}
  \frac{\partial\mathcal{C}}{\partial b_j^\ell} = \frac{\partial\mathcal{C}}{\partial z_j^\ell}\frac{\partial z_j^\ell}{\partial b_j^\ell} = \frac{\partial\mathcal{C}}{\partial z_j^\ell}.
  \label{eq:backprop_wrt_bias}
\end{equation}

We now seek an equation for ${\partial\mathcal{C}}/{\partial z_j^\ell}$. This quantity compares a neuron's pre-activation against the cost function and is used to make adjustments to the weights and biases in the network. We will continue with the convention that neurons with index $k$ feed into neurons with index $j$.

We can express the sensitivity in the output layer, $L$, by looking at how small or large the rate of change of the cost function is with respect to the weighted input $z_j^L$ (we could, in principle, look at the rate of change with respect to the full output of the neuron, but this involves an extra computational step for the activation function, and the weighted input contains all the information we need). Thus,
\begin{equation}
  \frac{\partial\mathcal{C}}{\partial z_j^L} = \frac{\partial\mathcal{C}}{\partial y_j^L}\frac{\partial y_j^L}{\partial z_j^L}=\frac{\partial\mathcal{C}}{\partial y_j^L} \mathcal{A}'(z_j^L).
  \label{eq:weighted_input_sensitivity_scalar} 
\end{equation}
If Eq.~(\ref{eq:weighted_input_sensitivity_scalar}) is small, then the neuron is nearly optimal. 

We then work top to bottom through the neural network (hence, backpropagation) in order to determine the optimality of each neuron. For the next layer, $L-1$, the sensitivity of the cost function with respect to the neuron pre-activations is
\begin{equation}
  \frac{\partial\mathcal{C}}{\partial z_k^{L-1}} = \sum_{j=1}^{n_L} \frac{\partial\mathcal{C}}{\partial z_j^{L}} \frac{\partial z_j^L}{\partial z_k^{L-1}},
  \label{eq:sensitivity_L-1}
\end{equation}
where
\begin{equation}
  z_j^L = \sum_{k=1}^{n_{L-1}} w_{jk}^L \mathcal{A}(z_k^{L-1}) + b_j^L.
  \label{eq:preactivation_L}
\end{equation}
Using Eq.~(\ref{eq:preactivation_L}) in Eq.~(\ref{eq:sensitivity_L-1}) gives the sensitivity of the cost function with respect to a neuron's pre-activation in the penultimate layer
\begin{equation}
  \frac{\partial\mathcal{C}}{\partial z_k^{L-1}} = \sum_{j=1}^{n_L} \frac{\partial\mathcal{C}}{\partial z_j^{L}} w_{jk}^L \mathcal{A}'(z_k^{L-1}). 
  \label{eq:sensitivity_L-1_full}
\end{equation}
Then, in any arbitrary layer $\ell-1$, $\ell\in\{1, \cdots, L-2\}$, the sensitivity of the cost function with respect to any arbitrary neuron in that layer is
\begin{equation}
  \frac{\partial\mathcal{C}}{\partial z_k^{\ell-1}} = \sum_{j=1}^{n_{\ell}} \frac{\partial\mathcal{C}}{\partial z_j^{\ell}} w_{jk}^\ell \mathcal{A}'(z_k^{\ell-1}).
  \label{eq:sensitivity_general}
\end{equation}

The parameter updates in Eqs. (\ref{eq:backprop_wrt_bias}) and (\ref{eq:backprop_wrt_weight}) are obtained by a possibly very lengthy chain rule. In machine learning, we use \emph{automatic differentiation} to compute these quantities \footnote{Automatic differentiation is explored in literature such as Baydin, \emph{et al.} \cite{baydin} and Griewank and Walther \cite{griewank}}.  


\clearpage
\bibliography{bib}

\end{document}